\documentclass[10pt,journal,compsoc]{IEEEtran}
\usepackage{multirow}
\usepackage{color}
\newtheorem{definition}{Definition}
\newtheorem{assumption}{Assumption}
\newtheorem{proposition}{Proposition}

\ifCLASSOPTIONcompsoc
  \usepackage[nocompress]{cite}
\else
  \usepackage{cite}
\fi
\ifCLASSINFOpdf
   \usepackage[pdftex]{graphicx}
\else
\fi
\usepackage{amsmath}
\usepackage{amsfonts}

\usepackage[ruled]{algorithm2e}

\usepackage[caption=false,font=footnotesize,labelfont=sf,textfont=sf]{subfig}
\begin{document}
%
\title{Boosting Model Inversion Attacks with Adversarial Examples}
%
%
\author{Shuai~Zhou, 
        Tianqing~Zhu*, 
        Dayong~Ye, 
        Xin Yu,
        and~Wanlei~Zhou
\IEEEcompsocitemizethanks{\IEEEcompsocthanksitem Shuai Zhou, Dayong Ye, and Tianqing Zhu are with the Centre of Cyber Security and Privacy and the School of Computer Science, University of Technology Sydney, Ultimo, NSW 2007, Australia. *Tianqing Zhu is the corresponding author. 
E-mail: shuai.zhou@student.uts.edu.au, \{dayong.ye,~tianqing.zhu\}@uts.edu.au.
\IEEEcompsocthanksitem Xin Yu is with the School of Computer Science, University of Queensland, QLD 4072, Australia. E-mail: xin.yu@uq.edu.au.
\IEEEcompsocthanksitem Wanlei Zhou is with the Institute of Data Science, City University of Macau,
Macau, China. E-mail: wlzhou@cityu.mo.}
\thanks{Manuscript received Dec 31, 2022.}}

\markboth{Journal of \LaTeX\ Class Files,~Vol.~14, No.~8, August~2015}%
{Shuai \MakeLowercase{\textit{et al.}}: Boosting Model Inversion Attacks with Adversarial Examples}

\IEEEtitleabstractindextext{%
\begin{abstract}
Model inversion attacks involve reconstructing the training data of a target model, which raises serious privacy concerns for machine learning models. However, these attacks, especially learning-based methods, are likely to suffer from low attack accuracy, i.e., low classification accuracy of these reconstructed data by machine learning classifiers. Recent studies showed an alternative strategy of model inversion attacks, GAN-based optimization, can improve the attack accuracy effectively. However, these series of GAN-based attacks reconstruct only class-representative training data for a class, whereas learning-based attacks can reconstruct diverse data for different training data in each class. Hence, in this paper, we propose a new training paradigm for a learning-based model inversion attack that can achieve higher attack accuracy in a black-box setting. First, we regularize the training process of the attack model with an added semantic loss function and, second, we inject adversarial examples into the training data to increase the diversity of the class-related parts (i.e., he essential features for classification tasks) in training data. This scheme guides the attack model to pay more attention to the class-related parts of the original data during the data reconstruction process. The experimental results show that our method greatly boosts the performance of existing learning-based model inversion attacks. Even when no extra queries to the target model are allowed, the approach can still improve the attack accuracy of reconstructed data. This new attack shows that the severity of the threat from learning-based model inversion adversaries is underestimated and more robust defenses are required.  
\end{abstract}

}

\maketitle

\IEEEdisplaynontitleabstractindextext

\IEEEpeerreviewmaketitle

\IEEEraisesectionheading{\section{Introduction}\label{sec:introduction}}

\IEEEPARstart{M}{achine} learning as a service (MLaaS) is becoming a popular paradigm for providing online services to users, as is evidenced by the success of services like Amazon Rekognition~\cite{Amazon} and Google's cloud vision API~\cite{Google}. Unfortunately, public access to these trained models can give rise to privacy concerns, such as model inversion attacks~\cite{nips/WangFLKZM21,icml/StruppekHCAK22,GameAPT}. These attacks occur when an attacker repeatedly queries a model, allowing them to reconstruct the original training data and potentially exposing sensitive information or allowing unauthorized model training. The importance of understanding and defending against model inversion attacks cannot be overstated. The potential consequences of such attacks range from privacy violations to financial losses. Thus, the aim of this paper is to explore the actual capabilities of attackers who carry out model inversion attacks, and to develop improved defense strategies against such attacks.
 
\subsection{Assessing Attack Performance}
In this paper, we focus on the model inversion attacks towards image classification models. In this context, attackers' capabilities can be evaluated by the the classification accuracy of those reconstructed data by the victim model or other machine learning models, known as attack accuracy~\cite{cvpr/ZhangJP0LS20,iccv/ChenKJQ21,kahla2022label}. A higher classification accuracy of the reconstructed data indicates a higher degree of model inversion attack accuracy and success. Model inversion attacks can be executed through either an optimization-based or a learning-based strategy. Optimization-based methods mostly rely on a generative adversarial network (GAN) as a prior to produce realistic training data~\cite{cvpr/ZhangJP0LS20,iccv/ChenKJQ21,icml/StruppekHCAK22,Mirror2022}. GAN-based optimization can often easily synthesize machine-recognizable data for a targeted class. 
However, GAN-based attacks typically generate only few representative data for each class and require white-box access to the target model, which is impractical in the real-world~\cite{cvpr/ZhangJP0LS20,iccv/ChenKJQ21}. Learning-based attack methods, however, can produce a reconstruction for each individual training data and require only black-box access to the model, which is considerably more feasible~\cite{ccs/YangZCL19}. With this method, attackers train a new attack model, called an inversion model, that swaps the input and output of the target model. This paper will focus on enhancing the learning-based model inversion attacks.

\begin{figure}
    \centering
    \includegraphics[width=0.98\columnwidth]{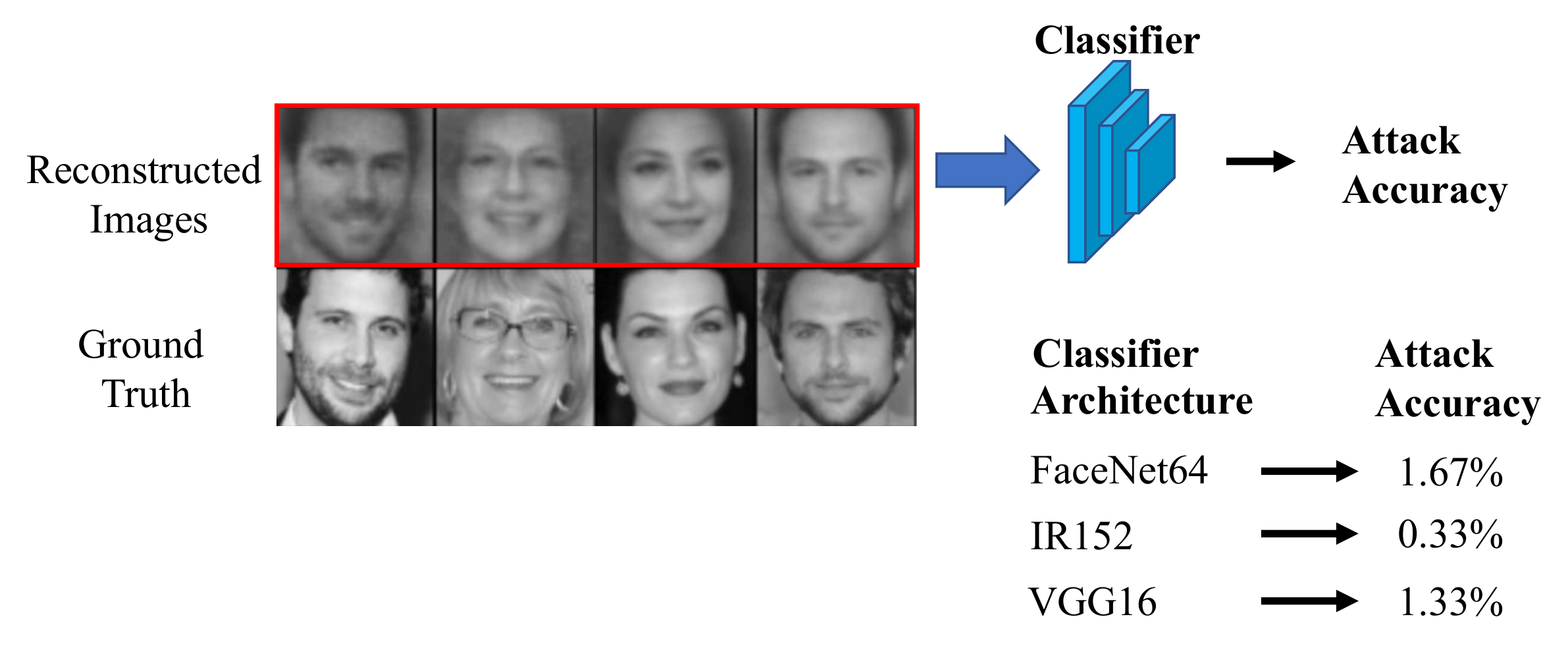}
    \caption{A case study which shows the low attack accuracy of the learning-based model inversion attacks.}
    \label{fig:case}
\end{figure}

In previous research on learning-based model inversion attacks, the classification accuracy of reconstructed data (i.e., attack accuracy) is not high. It means that these reconstructed data, such as the reconstructed face images, can only be used to disclose identities, rather than being correctly classified by the machine learning models. These previous attacks are almost impossible to reuse the reconstructed data for downstream malicious tasks, such as bypassing facial recognition. For instance, some reconstructed images are presented in Fig.~\ref{fig:case}, which are cited from~\cite{ccs/YangZCL19}. As shown in Fig.~\ref{fig:case}, despite the high quality of the reconstructed images, Kahla et al.~\cite{kahla2022label} reported that the attack accuracy of the learning-based method~\cite{ccs/YangZCL19} is still less than $2\%$. Therefore, researchers believed that the threat of a learning-based model inversion attack is less severe than the optimization-based counterpart. However, we argue that the severity of the threat from learning-based adversaries is underestimated and our paper shows that we can build a novel inversion attack with high attack accuracy. This finding underscores the increased danger posed by such attacks and highlights the need for corresponding defense measures. Our research questions can be subdivided into the following two sub-questions: What causes low attack accuracy of the learning-based model inversion attacks and how to improve the attack accuracy?

\subsection{RQ1: What Causes Low Attack Accuracy?}
Inspired by recent studies on adversarial examples~\cite{SzegedyZSBEGF13,DBLP:journals/corr/GoodfellowSS14}, we investigate the limitations of learning-based attacks in achieving high attack accuracy. Many existing learning-based attacks employ a loss function based on pixel-wise differences between the input image and the reconstructed image, such as the mean squared error (MSE)~\cite{ccs/YangZCL19}. We argue that the MSE-trained inversion model focuses only on the pixel-wise differences between the original data and the reconstructed data while overlooking to data's crucial semantic information (i.e., the essential features for classification tasks) and, therefore, cannot reconstruct data with high attack accuracy. This is due to the fact that even a negligible pixel-wise difference between them, say in the form of minor noise, may lead to a mis-classification. For example, the adversarial sample may lead to mis-classification with minor noise~\cite{SzegedyZSBEGF13}. 

Further, Yang et al.~\cite{nips/YangZZTT21} showed that the image data can be divided into class-related and class-redundant part. Building on their work~\cite{nips/YangZZTT21}, we also use the term ``class-related part'' to represent the essential information in an image for classification and ``class-redundant part'' to denote the remaining information that has less impact on the classification compared to the class-related part. We observed that in the model inversion process of reconstructing image data, the class-related part in image data is more likely to be omitted than the class-redundant part due to the minor variances in the pixels. This forces the trained inversion model to prioritize the recovery of the class-redundant part to decrease the training loss more quickly, even though the classification accuracy of the reconstructed data are primarily determined by the class-related part. Consequently, the reconstruction error of class-related part in the reconstructed data may be greater than that of class-redundant part, resulting in low attack accuracy. 
In other words, the class-related part in the reconstructed data are poorly reproduced in the data reconstruction, making it difficult for machine learning models to correctly classify the reconstructed data.

\subsection{RQ2: How To Improve Attack Accuracy?}
In this paper, to demonstrate we can improve the attack accuracy of model inversion attacks, we present a regularized loss function and an adversarial augmentation technique to address the limitations of MSE and improve the contribution of class-related part in the training process, respectively. 
We address the limitations of MSE by devising a new loss function that includes a penalty term for the semantic reproduction loss. This involves computing the cross-entropy loss between the predictions of the reconstructed data and the labels of the original data, which guides the inversion model to consider both the pixel-wise differences between the original and reconstructed data and the attack accuracy from model's perspective. This approach prevents the inversion model from overfitting to data's class-redundant part.  

In addition, to stop the inversion model from overlooking the class-related part, and to further improve the attack accuracy, an adversarial augmentation technique is introduced to increase the variety of data (particularly the class-related parts). 
This technique is based on the observation that adversarial perturbations are allocated to the areas of the image that neural networks primarily attend to, leading to misclassification~\cite{nips/YangZZTT21}. By generating adversarial examples via black-box adversarial attacks against the target victim model, we can locate the class-related part that determines the essential information relating to classification tasks in the training data. 
These class-related parts are the features that should be prioritized during the inversion model's training in order to increase the attack accuracy. These adversarial examples with their adversarial labels (i.e., incorrect labels predicted by the target model) are then injected into the training data as augmented data. The pairs of perturbed class-related part (in adversarial examples) and the corresponding adversarial labels can provide knowledge about the data's essential information relating classification tasks during the training of the inversion model. Thereby, the inversion model's ability to produce machine-recognizable reconstructed data that can be correctly classified with high probability by machine learning models is enhanced further.

\subsection{Contributions}
Overall, our contributions are summarized as follows:
\begin{itemize}

\item To the best of our knowledge, this is the first work to analyze the reasons behind the low attack accuracy of learning-based model inversion attacks. Specifically, existing learning-based methods typically produce some reconstructed images that cannot be correctly classified by the machine. We investigate this issue from both a theoretical and an experimental perspective.
\item This is the first study to utilize adversarial examples in exploring the privacy threats posed by model inversion attacks. Adversarial examples were originally proposed for investigating machine learning model security issues, whereas model inversion attacks primarily focus on privacy concerns. Traditionally, security and privacy problems have been investigated separately. However, this work suggests a potential direction for comprehensively investigating the privacy and security risks of deep learning models.
\item We introduce a novel definition of $k$-pseudo-label to guide the generation of adversarial examples for out-of-distribution (OOD) data. Generating adversarial examples typically requires having access to the correct labels for the data. However, in the case of out-of-distribution (OOD) data, valid labels do not exist, which poses a significant challenge for generating adversarial examples. Our proposed approach, using a $k$-pseudo-label, effectively addresses this issue and enables the generation of adversarial examples even for OOD data.
\item In the past, learning-based model inversion attacks were generally considered to have low attack accuracy. However, in this paper, we propose a new training paradigm for learning-based model inversion attacks that allows for the recovery of private data with a high level of attack accuracy. This more accurately assesses the true threat of learning-based model inversion attacks and helps to improve defense strategies against such attacks.

\end{itemize}

\section{Related Work}
\subsection{Model Inversion Attacks}
\subsubsection{Optimization-based Model Inversion Attacks}
Model inversion attacks can be divided into two classes: optimization-based attacks and learning-based attacks. Given a target victim model, the objectives of optimization-based attacks are usually to reconstruct one unknown data that is used in the training of the target model. Fredrikson et al.~\cite{ccs/FredriksonJR15} proposed the first optimization-based model inversion attacks against shallow neural networks based on a gradient descent strategy. For deeper networks, GAN is usually used to restrict the optimization space during data reconstruction. Zhang et al.~\cite{cvpr/ZhangJP0LS20} introduced a GAN to restrict the optimization space in the data reconstruction, which can be considered the distribution prior and guide the inversion process. Chen et al.~\cite{iccv/ChenKJQ21} designed a special GAN to distill knowledge about the private data from the target model and further improve the attack performance. Built upon this approach, Kahla et al.~\cite{kahla2022label} designed a label-only attack in the practical setting where adversaries can only obtain the hard labels. Similarly, Zhu et al.~\cite{zhu2022label} also developed a label-only attack to reconstruct a class-representative of training data. Wang et al.~\cite{nips/WangFLKZM21} formulated the model inversion attack as a variational inference problem, proposing a practical variational objective that is optimized by combining a GAN and a deep flow model.

However, the GAN trained for a specific target model in these works makes the model inversion attacks time-consuming and non-robust against the distribution shifts between datasets. Therefore, Struppek et al.~\cite{icml/StruppekHCAK22} proposed a flexible and robust attack that involves training a single GAN to attack multiple target models. Notably, GAN-based optimization strategies usually require white-box access to the target model. Mirror~\cite{Mirror2022} leverages a StyleGAN that was pre-trained on public data to reconstruct private data in a black-box setting, although it might not always be practical to obtain a pre-trained GAN in all tasks. 
\subsubsection{Learning-based Model Inversion Attacks}
In general, optimization-based model inversion attacks require either white-box access to the target model, or a GAN (pre-)trained on public data. In addition, these optimization-based attacks typically produce a small number of class-representative data for each class, as opposed to multiple different reconstructed data for various original data within one class. In contrast, Yang et al.~\cite{ccs/YangZCL19} presented the first seminal learning-based model inversion attack, where a separate model is trained to reconstruct the private input data from predictions made by the target model. Their method can achieve accurate inversion of individual private data in adversarial settings instead of class-representative inversions, and has increased our awareness of the necessity to protect privacy. In the real world, learning-based attacks are more feasible than optimization-based attacks. However, these reconstructed data by learning-based model inversion attacks have been empirically shown to be hard for other classifiers to recognize~\cite{kahla2022label}, despite resembling the original private data for human perception. 

Zhao et al.~\cite{iccv/ZhaoZXL21} focused on a different case where the target task (i.e., the victim model's task) was distinct from the attack task. For instance, as the attack task, they aimed to reconstruct the facial image of users to steal their identity, while the victim model's target task was to predict an emotion. By their thinking, the target model's explanations increased the privacy risks of the model inversion attacks. Hence, they developed a multi-modal transposed CNN architecture to reconstruct the private data by introducing these explanations. This worked to improve the inversion performance much better than using prediction alone, as well as the classification accuracy of the reconstructed data when fed to other classifiers. This method, however, cannot be adapted to scenarios where both target task and attack task share the same goal, like identity prediction, as is the case with Yang et al.'s work~\cite{ccs/YangZCL19}. 

\subsection{Adversarial Examples}
\subsubsection{White-box adversarial attacks}
Adversarial attacks demonstrate that neural networks are not robust to maliciously crafted small noise. The perturbed data, i.e., the adversarial examples, are likely to be misclassified by the target model~\cite{SzegedyZSBEGF13,iclr/MadryMSTV18}. For example, the search for the smallest adversarial perturbation can be formalized as an optimization problem and solved by L-BFGS~\cite{SzegedyZSBEGF13}. Alternatively, the adversarial perturbation can be effectively computed by calculating the gradient of the loss function with respect to the input of the target model~\cite{DBLP:journals/corr/GoodfellowSS14}. However, most of them require white-box access to the target model, which is unrealistic in the real world. In the majority of cases, an adversary only has access to the output from the target model. For this reason, many black-box adversarial attacks are designed to produce effective adversarial examples from the target model's query responses~\cite{zhou2021adversarial}. 
\subsubsection{Black-box adversarial attacks}
Papernot et al.~\cite{PapernotBB}, for example, presented the first black-box adversarial attack by training a substitute model to emulate the target model. This type of attack relying on the ability of adversarial examples to generalize across models, also known as transfer-based attacks, usually requires access to the training data distribution of the target model. Other methods try to estimate the target model's gradients~\cite{ZOO,AutoZOOM} when the target model is not accessed or exploit greedy local-search techniques to implement gradient-free adversarial attacks~\cite{squareattack}. Guo et al.~\cite{icml/GuoGYWW19} is one team that provided an intriguingly simple strategy resulting in highly efficient performance. They use the output probabilities as the proxy for the distance to the decision boundary to guide the generation of adversarial images. By adding (or subtracting) a randomly chosen vector to the target image and feeding it to the target model iteratively, the attacker can check whether or not the vector is pointing towards the decision boundary.

\subsection{Discussion}
When compared to the existing works, our method has some advantages as follows. 
\begin{itemize}
    \item \textbf{Diverse reconstructions.} The majority of model inversion attacks employ an optimization-based strategy and only reconstruct class-representative training data for each class~\cite{iccv/ChenKJQ21,cvpr/ZhangJP0LS20,Mirror2022}. Similar to Yang et al.'s work~\cite{ccs/YangZCL19}, our method focuses on the learning-based model inversion attack, which can produce diverse reconstructions for distinct original data within a class. 
    \item \textbf{More practical.}  The only existing method that specifically addresses the problem of low attack accuracy for learning-based model inversion attacks is the exploiting explanations method proposed by Zhao et al.~\cite{iccv/ZhaoZXL21}. This method, however, requires the introduction of an additional explanation model for the target victim model and is designed for a special scenario, which makes it impractical for typical learning-based model inversion attacks~\cite{ccs/YangZCL19}. In contrast, our strategy is tailored to the same model inversion scenarios as the seminal work by Yang et al.~\cite{ccs/YangZCL19}, and it requires no additional information or training data, making it more practical for real-world use.
    \item \textbf{Different from adversarial training.} In addition, our adversarial augmentation method, which augments the inversion model's training data with adversarial examples, superficially resembles adversarial training on classification tasks~\cite{iclr/MadryMSTV18}. However, the underlying ideas are completely different. In adversarial training, the classifier is trained on the adversarial examples with correct labels in order to increase this classifier's robustness against noises. In contrast, our adversarial augmentation assigns adversarial labels (i.e., the incorrectly predicted labels) to the adversarial examples, which forces the inversion model to be more sensitive to the perturbations mainly lying in the class-related part.
\end{itemize}

\section{Preliminaries}
\label{sec:pre}
\subsection{Learning-based Model Inversion Attacks}
Machine learning models providing online services has been a popular paradigm~\cite{Google, cloud, zhang2022defend}, however, Yang et al.~\cite{ccs/YangZCL19} find that the training data of publicly accessible model can be reconstructed through learning-based methods, where the learning process is to find an optimal inversion model $I$ which can minimize the following objective: 
$$\mathcal{L}(I) = \mathbb{E}_{x\in X}\mathcal{R}(x, I(T(x)))$$
where $\mathcal{R}$ represents the metric measuring the reconstruction error between reconstructed data $I(T(x))$ and the original data $x$, e.g., the mean squared error (MSE) in the image domain. When $R(x, I(T(x)))$ is extremely small, the reconstructed data $I(T(x))$ can be similar to the original data $x$. However, it is still hard to correctly classify the reconstructed data with a machine learning model, even though humans can do it quite easily~\cite{kahla2022label}. So, although the reconstructed data is similar to the original data, the machine learning model classifies the data into a different classes. That is, the predicted label of $x$ and the predicted label of $I(T(x))$ are different: $$\arg\max_i T^c_i(x)\neq \arg\max_j T^c_j(I(T(x))),$$ where $T^c$ is a classifier with same data distribution as $T$, and $T_i(x)$ represents the $i$-th element of output $T(x)$.  

\subsection{Black-box adversarial attacks: SimBA}
Let $T$ be the target model such that $\arg\max_i T_i(x)$ equals $y$ with a high probability, where $x$ is the input data, and $y$ is the ground truth label of $x$. Adversarial attacks are designed to seek a perturbation $\delta$ to achieve $\arg\max_i T_i(x+\delta)=y'$, where $y'\neq y$. The norm of perturbation $\delta$ is typically bounded by a small scalar $\eta$, which makes the perturbed data $x+\delta$ indistinguishable from the clean data $x$. More concisely, the adversary in adversarial attacks aims to seek a perturbation $\delta$ that satisfies the following:
$$\arg\max_i ~T_i(x+\delta) = y' \quad{\rm s.t.}\quad y'\neq y, ~ \left\|\delta\right\|\leq \eta.$$
When the output $y'$ of the perturbed data $x+\delta$ is specified by attackers, this type of attacks is referred to as targeted adversarial attacks. In contrast, the purpose of the untargeted attack is to mislead the target model into producing a different output than $y$, regardless of whatever output it is. 

When the attacker has only black-box access to the target model, it will be challenging to craft effective adversarial examples.  SimBA~\cite{icml/GuoGYWW19}, a simple black-box attack method, was proposed to solve this issue by repeatedly querying the target model. The attacker attempts to move the clean data towards the decision boundary by continually perturbing the data until the data crosses the decision boundary. The distance between data $x$ and the decision boundary is approximated by the output probabilities $T_i(x)$ of data $x$. In particular, during each iteration, the attacker checks whether the probabilities decrease after adding (or subtracting) the random perturbations $\delta=\epsilon q$, where $\epsilon$ is a scalar indicating the perturbation size in each iteration, and $q$ is the direction of the perturbation. The reduction in probabilities indicates that the perturbed data are closer to the decision boundary than the unperturbed data; consequently, the data will be updated by adding (or subtracting) the selected perturbation. The candidate direction vectors should be orthonormal so that the directions $q$ from different iterations do not cancel each other out. Using the untargeted attack as an illustration, the iteration will end when the label of the perturbed data, $y' = \arg\max_i ~T_i(x+\delta)$, differs from the ground truth label $y$ of clean data $x$.

\section{Reasons for poor attack accuracy}
We investigate the reasons why existing black-box model inversion attacks tend to achieve poor attack accuracy in this section. As mentioned in Section~\ref{sec:pre}, $\arg\max_i T^c_i(x)\neq \arg\max_j T^c_j(I(T(x)))$, even though the distortion between the original and reconstructed data $R(x, I(T(x)))$ is extremely small. Inspired by Yang et al.'s recent study~\cite{nips/YangZZTT21} on class-disentanglement tasks, we argue that this phenomenon might be caused by two factors - an underdeveloped loss function and overlooking the class-related part.

\subsection{Underdeveloped loss function.}
Most existing methods include a loss function based on the difference between the original training image and the reconstructed image, e.g., the mean squared error (MSE) of two images. As a result, the well-trained inversion model is able to reconstruct an image with a small MSE difference from the original image. Despite the fact that an extremely small difference indicates that a reconstructed image is extremely similar to the original one, this does not guarantee that the semantic information of the two images will be identical. Adversarial examples provide some evidence of this phenomenon~\cite{iclr/MadryMSTV18}. Specifically, the MSE difference between a clean image and its adversarial example is typically small. Indeed, these imperceptible distortions in an adversarial example can mislead a model into predicting a wrong class for this adversarial example. As such, the conventional loss functions, like MSE, used in existing learning-based model inversion attacks tend not to be good enough to capture the real distribution of the original data. Hence, the reconstructed data might still be incorrectly classified even if there are only small differences between the two sets of data. So, the underdeveloped loss function is insufficient to guarantee the accurate reconstruction of crucial information determining classification tasks in the original data.

\begin{figure}
    \centering
    \includegraphics[width=0.9\columnwidth]{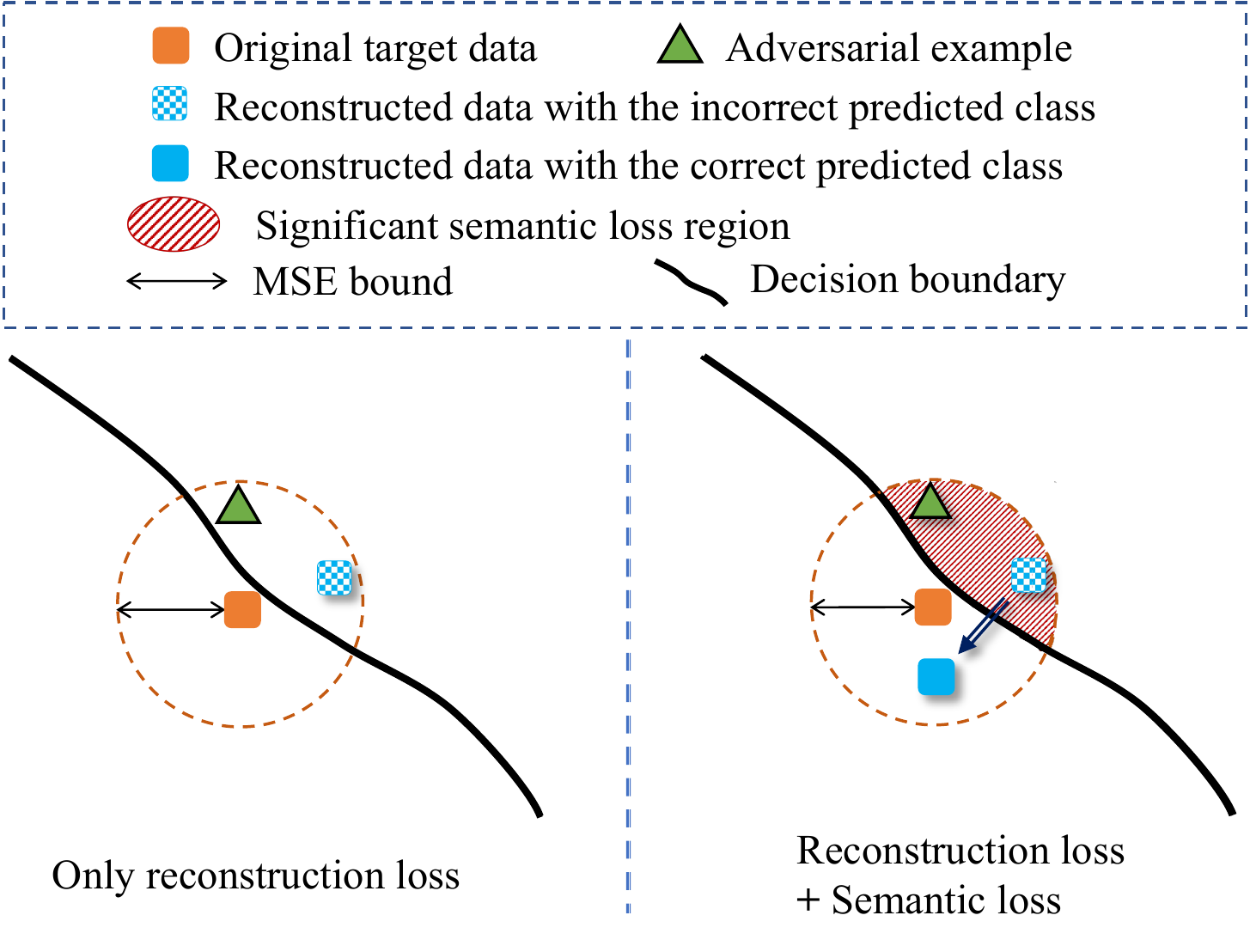}
    \caption{The intuitive explanation for the poor attack accuracy in the existing model inversion attacks. In the left part, when only the reconstruction loss is used to train the inversion model, the reconstructed data may cross the decision boundary. Like adversarial examples, these reconstructed data are close to the original target data but have a different predicted class, resulting in low attack accuracy.  In the right part, a semantic loss is incorporated with reconstruction loss, and the reconstructed data from the opposite side of the decision boundary will result in a significant semantic loss. Consequently, the inversion model will be forced to produce reconstructed data that is in the same side as the original data.}
    \label{fig:reason}
\end{figure}

We present an example in the left part of Fig.~\ref{fig:reason}. If only the MSE (referred to as reconstruction loss) is used as the loss function, the inversion model is trained to produce an example that is close to the original data in terms of MSE. The area enclosed by the red dashed circle represents the region that are sufficiently close to the original data in terms of MSE distance, and therefore, the data reconstructed by the inversion model that is trained on only reconstruction loss are drawn from this region. However, if the original target data that the attacker aims to reconstruct is close to the decision boundary, this region of potential reconstruction would cross the decision boundary. Consequently, the reconstructed data might be located on the opposite side of the decision boundary, similar to adversarial examples, and $\arg\max_i T_i(x)\neq \arg\max_j T_j(I(T(x)))$. When the semantic loss (which will be introduced in Section.~\ref{sec:loss}) is incorporated with the reconstruction loss (as depicted in the right part in Fig.~\ref{fig:reason}), the reconstructed data from the opposite side of the decision boundary will result in a significant semantic loss. As a result, the likelihood of the inversion model producing data from the region on the opposite side of the decision boundary (i.e., significant semantic loss region in Fig.~\ref{fig:reason}) decreases greatly. It will be more likely that the inversion generates reconstructed data with the correct predicted class.

\subsection{Overlooked class-related information.}
\begin{figure}[t]
\centering
\includegraphics[width=0.98\columnwidth]{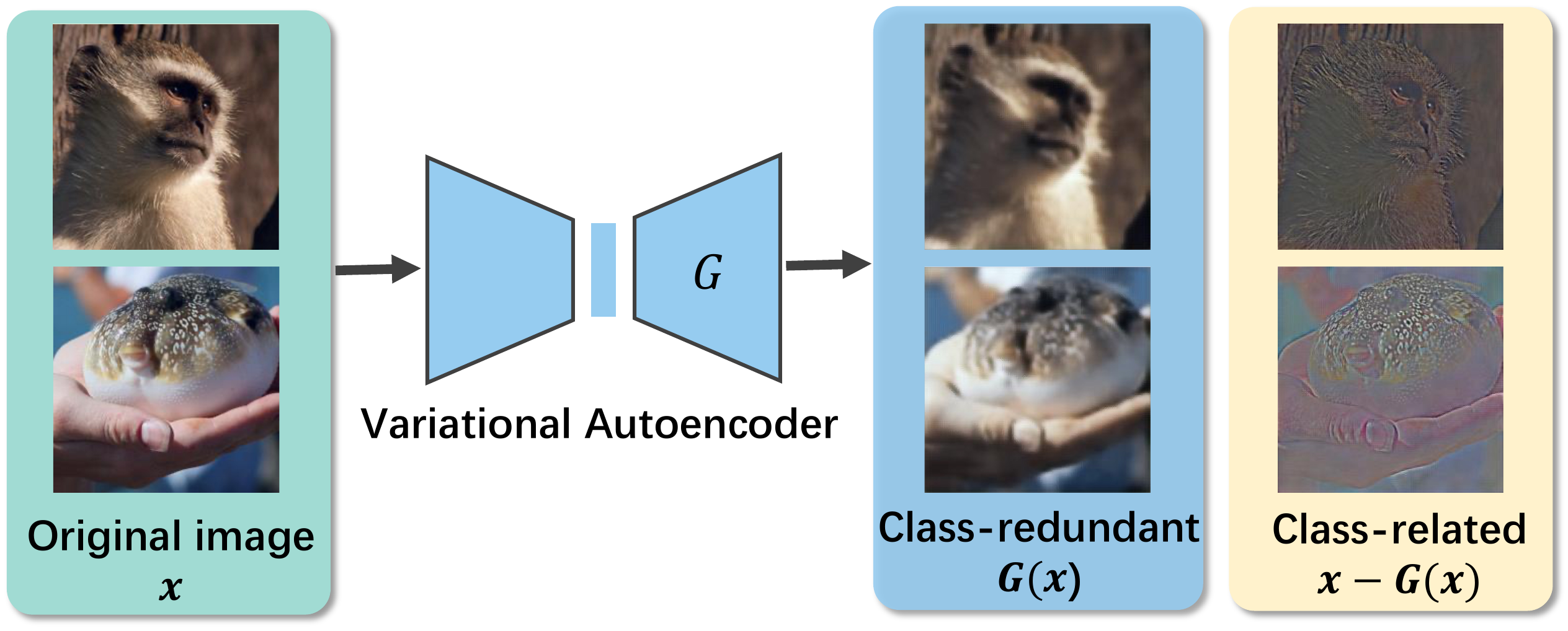} 
\caption{As shown in the previous work~\cite{nips/YangZZTT21}, images $x$ are disentangled by a variational autoencoder, where $G(x)$ and $x-G(x)$ represent the class-redundant and class-related information.}
\label{fig:CDVAE}
\end{figure}

We also observed that class-related part in an image tends to be ignored during image reconstruction process based solely on differences between data, because class-redundant information contributes more to the reconstruction loss. More specifically, Yang et al.~\cite{nips/YangZZTT21} disentangled class-related and class-redundant parts in images by training a variational autoencoder. Their results are shown in Fig.~\ref{fig:CDVAE}. What the figure shows is that the class-redundant part $G(x)$ tends to be colorful with a larger variance in pixels, while the class-related part $x-G(x)$ is relatively monotonous. 

The reconstruction of one image can be viewed as a summation of the respective reconstructions of the two separate parts, class-redundant $G(x)$ and class-related $x-G(x)$. The objective of the reconstruction is to minimize the difference between the original images (e.g., $G(x)$ or $x-G(x)$) and their reconstructed images. Assuming that the pixels in $G(x)$ follow a Gaussian distribution $N(\mu_1, \sigma_1)$, and pixels in $x-G(x)$ follow another Gaussian distribution $N(\mu_2, \sigma_2)$, it will always hold that $\sigma_1>\sigma_2$ because most pixels in the class-related information $x-G(x)$ are similar to each other and approximately black. Noting that it is easy for a neural network to identify the distribution of pixels, although determining their exact values is hard, which is the ultimate goal of learning an inversion model. Hence, we can assume that the inversion model can easily generate an initial random image with pixels following $\mathcal{N}(\mu_i, \sigma_i)$ (the pixels' distribution of the original image). To this end, the training loss that needs to be minimized in the inversion model's training process can be computed as follows (using MSE as the example):
\begin{equation}
\begin{split}
	{\rm MSE}(x^i,\hat{x}^i) &= \sum_k^K(x^i_k-\hat{x}^i_k)^2\\
	&\approx K\cdot\mathbb{E}_{\{x^i_k, \hat{x}^i_k \in \mathcal{N}(\mu_i, \sigma_i)\}}^2\left[x^i_k-\hat{x}^i_k\right] \\
	&~~+K\cdot\mathbb{D}_{\{x^i_k, \hat{x}^i_k \in \mathcal{N}(\mu_i, \sigma_i)\}}(x^i_k-\hat{x}^i_k)\\
	&= 0 + 2K\sigma_i^2\\
	&=2K\sigma_i^2
\end{split}
\end{equation}
where $x^i_k$ is the $k$-th pixel in the original image components $x^i$ (i.e., $G(x)$ or $x-G(x)$), $K$ is the total number of pixels, and $\hat{x}^i_k$ represents the $k$-th pixel in the reconstructed images $\hat{x}^i$. Both follow $\mathcal{N}(\mu_i, \sigma_i)$. Here, $\mathbb{E}$ denotes the expectation, and $\mathbb{D}$ is the variance. 

Therefore, the difference to be minimized (i.e., the training loss) is proportional to the variance $\sigma_i$ of original images. Compared to the class-related parts, the error of reconstructed class-redundant parts with a larger variance $\sigma_1$ contributes more to the loss. Thus, an accurate reconstruction of class-redundant parts can also result in a more obvious reduction of the training loss when compared to the class-related parts. This explains why class-related parts tends to be overlooked in learning-based model inversion attacks. For more details, see Section~\ref{sec:theo1}.

\section{Proposed Method}
\subsection{Problem Formulation}
\begin{figure*}[t]
\centering
\includegraphics[width=0.8\textwidth]{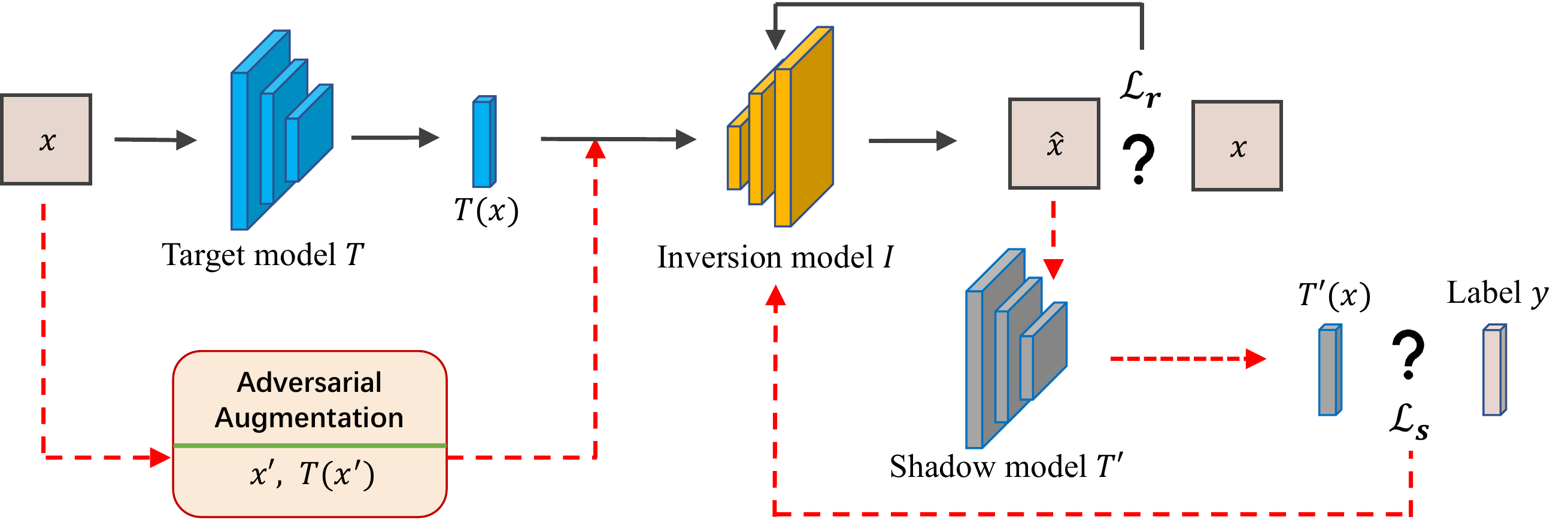} 
\caption{Workflow of a learning-based model inversion attack. The attacker trains an inversion model $I$ to reconstruct some private input $x$ from a prediction $T(x)$ returned by the target model $T$. The solid lines represent the data flow during training of the inversion model for existing methods. Based on that, our proposed training paradigm adds some new data flow. The dashed lines depict the supplementary data flow. First, $T'$ is a shadow model that was trained by querying $T$. The clean data $(x, T(x))$ and the augmented data $(x', T(x'))$ constitute the inversion model's training data. In addition to the reconstruction loss $\mathcal{L}_r$, a semantic loss $\mathcal{L}_s$ is also computed based on the shadow model $T'$. 
}
\label{fig:workflow}
\end{figure*}

In our setting, a black-box attacker can only query the target model and access the predictions, i.e., the soft labels of the target model; the internal information and the training data of the models are all unknown. However, attackers possess attack data from the same domain as the target model's training data. For example, if the target model is an identity classifier for $100$ persons, the attacker is assumed to have some face images not included in the target model's $100$ individuals. Fig.~\ref{fig:workflow} depicts the workflow diagram with solid lines. Given a target model $T$, the attacker will train an inversion model $I$, which swaps the input and output of the target model to reconstruct the original input $x$ from its predictions $T(x)$. 

Model inversion attacks are designed to acquire private data. 
And, for the attack to work, the reconstructed data $\hat{x}$ needs to be extremely similar to the original $x$. Following previous works~\cite{ccs/YangZCL19}, the reconstruction loss $\mathcal{L}_r$, which measures the pixel-wise differences between the original data and the reconstructed data, can be used to maximize the pixel-wise similarity of the original data and the reconstructed data.

Therefore, the goal in our model inversion attack is to enhance the attack accuracy (i.e., the classification accuracy of the reconstructed data by the machine learning classifiers). This can ensure the reconstructed data $\hat{x}$ are able to be re-used to the fullest extent possible, likely for malicious tasks such as using a reconstructed photo to bypass a face recognition security system. This goal requires that the reconstructed images contain information of class-related parts. To this end, our approach involves a regularized loss function and an adversarial augmentation technique. The regularized loss function, which contains the semantic loss $\mathcal{L}_s$, maximizes the model's ability to reconstruct machine-recognizable data. The adversarial augmentation procedure then injects adversarial examples with adversarial labels into the training data. This further emphasizes the class-related parts in the data reconstruction process. 

Based on the setting and goals, we introduce our new training paradigm for learning-based model inversion attacks, which was designed to ensure the reproduced data can be correctly classified by classifiers. To address the limitations caused by the underdeveloped loss function, we devised a regularized loss function with a penalty term for classifying the reconstructed data in our paradigm. In addition, our adversarial augmentation technique adds adversarial examples with the incorrect predicted labels into the inversion model's training data, which can capture the class-related parts in the images and contribute to emphasizing the class-related parts by increasing the variety. As depicted in Fig.~\ref{fig:workflow}, the dashed lines represent the additional data flow in our proposed training paradigm. First, the attacker trains a shadow model $T'$ using a knowledge distillation strategy to roughly mimic the target model $T$. Even though the train/test data or architecture is unknown, we found this distillation process can still be performed easily in black-box settings. After building the shadow model, a regularized loss function is used to train an inversion model, which considers both the data reconstruction loss and semantic loss simultaneously. Noting that the computation of semantic loss $\mathcal{L}_s$ is based on the shadow model $T'$. Moreover, the adversarial augmentation procedure produces adversarial examples against the target model $T$, and these adversarial examples are then inserted to the training data of the inversion model as augmented data. 

\subsection{The regularized loss function}
\label{sec:loss}
As shown in Fig.~\ref{fig:workflow}, the most common loss function used in existing learning-based inversion tasks is a reconstruction loss - for example, the average difference between the original and the reconstructed data, measured in terms of MSE:
\begin{equation}
	\mathcal{L}_r = \mathbb{E}_{x\in X}{\rm MSE}(x, \hat{x})
\end{equation} 
To design a proper loss function that considers the classification results of the reconstructed data, an additional penalty term is explicitly added to the loss - this being the semantic loss measuring the cross-entropy (CE) between the prediction of the reconstructed data and the label of the original data:
\begin{equation}
	\mathcal{L}_s = \mathbb{E}_{x\in X}\left[{\rm CE}(T(\hat{x}), y)\right]
\end{equation}
where $x$ is the original data with label $y$, and $\hat{x}$ is the reconstructed data. $T$ denotes the target model under attack. Given a normal sample, $y$ is the real label. However, given an augmented sample (which is introduced as the second technique of our training paradigm), $y$ is a label assigned by the adversary. As a result, our proposed loss function is formulated as follows:
\begin{equation}
\begin{split}
	\mathcal{L} &= \mathcal{L}_r + \lambda\cdot\mathcal{L}_s\\
	&=\mathbb{E}_{x\in X}\left[{\rm MSE}(x, \hat{x}) + \lambda\cdot{\rm CE}(T(\hat{x}), y)\right]
\end{split}
\end{equation}

Note that directly applying this loss function $\mathcal{L}$ during the inversion model's training is impractical due to two limitations, including black-box access to the target model and the out-of-distribution attack data. First, the adversary has only black-box access to the target model $T$. That means the loss can be computed by querying the target model. However, the gradient of the loss w.r.t. $\hat{x}$ is not accessible because the parameters of $T$ are unknown. Consequently, this regularized loss function with a semantic loss $\mathcal{L}_s$ cannot be directly applied in the backpropagation. To address this problem, a substitute for the target classifier is trained using a knowledge distillation strategy - this substitute being the shadow model. From the attack data and the query results returned by the target model, i.e., soft label, knowledge from the target model can be distilled into a shadow model that mimics the behavior of the target model. 

It is worth noting that training the shadow model does not require extra queries because the attack data for training the shadow model can also be used to train the inversion model. For example, if the attackers aim to recover some faces of unknown persons, it is enough to collect some pictures of celebrities available publicly on the internet, which can be used to train the shadow model and the inversion model sequentially. It is assumed that these publicly collected data of celebrities (i.e., attack data) do not overlap with the private data used to train the target model, and therefore, these attack data are out-of-distribution, bringing about the second limitation. In particular, the out-of-distribution data do not have valid ground truth labels as they do not belong to any of the classes of the target model. Due to the lack of ground truth labels, the semantic loss cannot be computed directly. In order to address this limitation, we replace the ground truth label $y$ with the output of the target model (also known as the soft label) when computing the semantic loss.

\subsection{Adversarial augmentation}

The adversarial augmentation is designed to boost the variety of class-related parts (i.e., the essential features relating classification tasks) in the training data. Adversarial examples are generated and then injected into the training data of the inversion model. The underlying idea is that adversarial perturbations inserted into the adversarial examples are more likely to be allocated to class-related parts than class-redundant parts~\cite{nips/YangZZTT21}, and the perturbed class-related parts ultimately lead to misclassification. Based on this observation, we believe adversarial samples can be used to capture the class-related parts that determine the classification results. These class-related parts are the features that should be prioritized during the inversion model's training. This inspired us to add adversarial examples of normal training data into the training set of the inversion model so as to increase the variety of the class-related parts in images. 
\subsubsection{Producing adversarial examples}
Due to the black-box access to the target model, we have to use black-box adversarial attacks to produce adversarial examples of attack data. There are various existing black-box attacks that do not need access to the structure and parameters of the target model~\cite{PapernotBB,squareattack,icml/GuoGYWW19}. In this paper, we choose Guo et al.'s method~\cite{icml/GuoGYWW19}. SimBA, owing to its simplicity and efficacy. A successful adversarial example will get a wrong predicted label that differs from the ground truth label. However, as these attack data are out-of-distribution, they lack valid ground truth labels, making it difficult to determine whether producing adversarial examples was successful or not. As a result, we propose pseudo-labels for the out-of-distribution attack data to guide the generation of adversarial examples. As a proxy of the real label, the pseudo-label should be resistant to non-malicious random noises. In other words, when the random noises are inserted into the clean inference data, the predicted pseudo-label should be consistent with that of the clean data. The pseudo-label with this property are defined based on the querying output of the target model as follows.

\begin{definition}($k$-pseudo-label)
\label{def:pse}
Let $T$ be a target model and $x$ be out-of-distribution data. If the output of target model $T$ on $x$ is $c = T(x)$, the $k$-pseudo-label of $x$ is an unordered group of classes with the highest $k$ probabilities in the output: $(l_1,l_2,\cdots,l_k)$. That means:$$c[l]\leq \min(c[l_1],c[l_2],\cdots,c[l_k]), \forall~ l\notin(l_1,l_2,\cdots,l_k).$$
\end{definition}

A wrong pseudo-label is required to have no intersections with the original pseudo-label. For example, a target model $T$ is designed to classify $10$ people. When another one people that is not included in $10$ people is fed into $T$, the querying output might be as follows:
$$[0.1,0.1,\textbf{0.2},\textbf{0.3},\textbf{0.2},0.02,0.02,0.02,0.02,0.02],$$
and its $3$-pseudo-label is $(2,3,4)$. A wrong $3$-pseudo-label leading to misclassification should have no overlap with this original $3$-pseudo-label $(2,3,4)$. For example, $(0,1,7)$ can be a valid wrong pseudo-label while $(0,1,4)$ is not a valid wrong pseudo-label.

More specifically, we adapted Guo et al.'s SimBA~\cite{icml/GuoGYWW19} to produce an adversarial example for each training sample. SimBA proposed to use the output probabilities for the proxy of distance to the decision boundary to guide the generation of adversarial examples. Likewise, we use the average of $k$ probabilities corresponding to $k$ classes in the pseudo-label as the proxy for distance to the decision boundary. The specific details are outlined in Algorithm~\ref{alg:simba}. 

This procedure takes as input the out-of-distribution attack data $x$, the noise added in each iteration $\epsilon$, and the number of classes in pseudo-labels $k$ (i.e., pseudo-label size). As a result, a perturbation $\delta$ is crafted, which will be added to the original data $x$ in order to transform $x$ into an adversarial example. In the initialization phase in  (line $1$-$3$), $\delta$ is initialized to $0$. And $pse = topk(T(x))$ represents the $k$-pseudo-label of $x$ as specified in Definition~\ref{def:pse}. Given a confidence score vector $T(x)$, the average of $k$ probabilities corresponding to $k$ classes in $pse$ is denoted as $Pr_k$, which serves as the proxy for distance to the decision boundary and guides the update of the perturbation $\delta$. At each iteration, a direction $q$ is randomly selected from the direction space $Q$. The data $x$ will be updated by adding the noise $x+\epsilon q$ (or subtracting the noise $x-\epsilon q$) when the addition (or subtraction) of the noise decreases the average probabilities of the pseudo-label $pse$. 

\begin{figure}
    \centering
    \includegraphics[width=0.65\columnwidth]{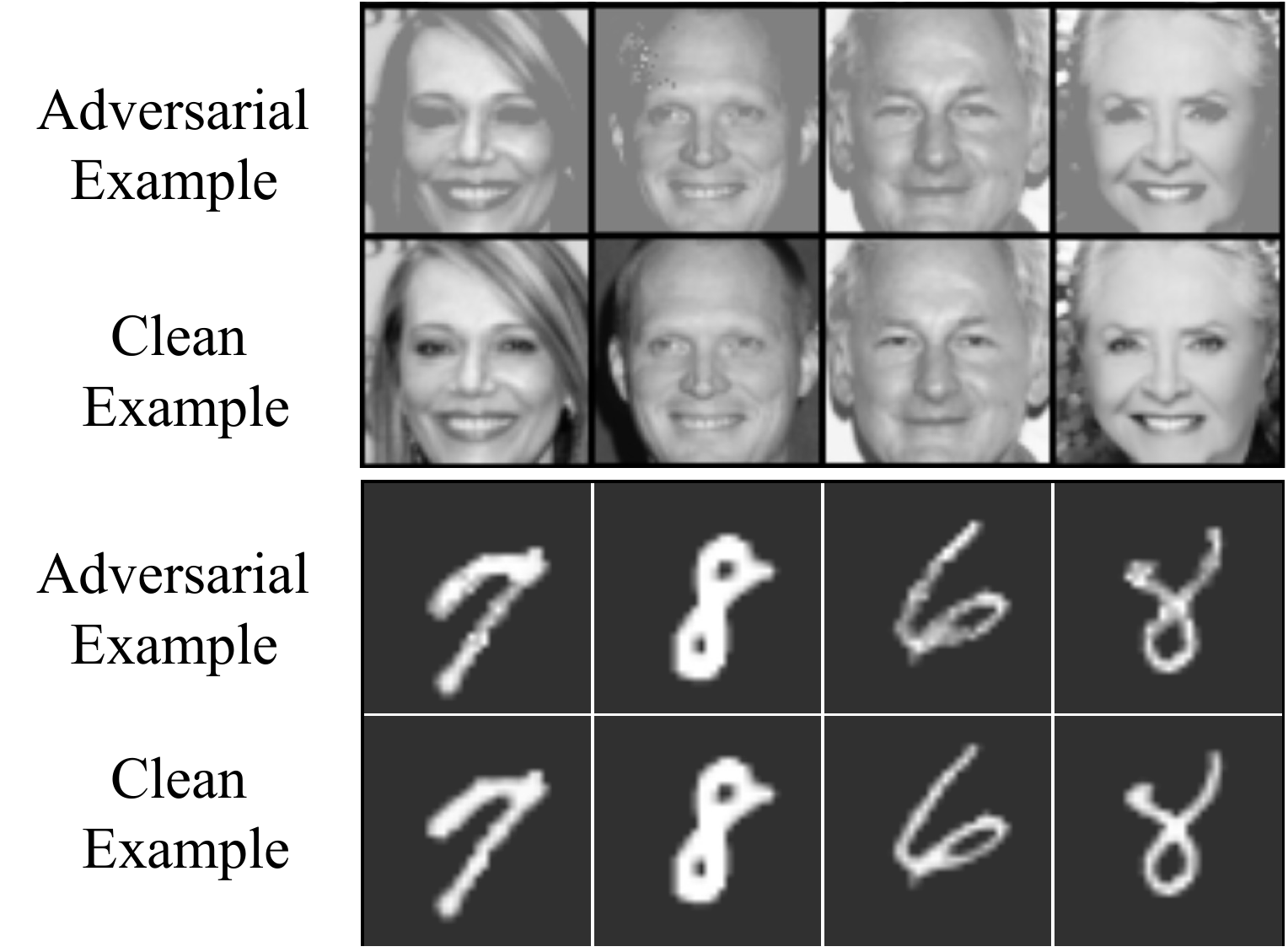}
    \caption{The adversarial examples produced on FaceScrub and MNIST by the adapted SimBA with pseudo-labels.}
    \label{fig:adv}
\end{figure}

In order to prevent the directions from different iterations from cancelling each other out, the direction space $Q$ consists of orthonormal candidate direction vectors. The perturbation $\delta$ is updated repeatedly until the pseudo-label is changed, that is, until the $k$ classes in updated $pse$ no longer overlap with the classes with the $k$ highest probabilities of the original $T(x)$. In other words, when the perturbed data $x+\delta$ is fed into the target model, if the output is a wrong pseudo-label that has no intersections with the original pseudo-label, this perturbed data can be considered a successful adversarial example. Fig.~\ref{fig:adv} shows the adversarial examples produced by our adapted SimBA with pseudo-labels. The distortion between the adversarial and clean examples is hard to distinguish, but the adversarial example can result in a different pseudo-label than the clean example.

\begin{algorithm}
    \caption{SimBA with pseudo-labels.}
    \label{alg:simba}
    \LinesNumbered 
    \SetAlgoLined
    \KwIn{The target model $T$\;\\ \qquad ~~~~Out-of-distribution data $x$\;\\ \qquad ~~~~Noise $\epsilon$\;\\ \qquad ~~~~Pseudo-label size $k$\;}
    \KwOut{Perturbation $\delta$\;}
    $\delta =0 $\; 
    $pse = top_k(T(x))$ is the $k$-pseudo-labels of $x$\;
    $Pr_k = \dfrac{\sum(T(x)[pse])}{k}$\;
    \While{$pse=top_k(T(x))$}{
    Pick randomly a direction $q\in Q$\;
    \For{$\alpha\in \{-\epsilon, \epsilon\}$}{
    $pse' = top_k(T(x+\delta+\alpha q))$\;
    $Pr_k' = \dfrac{\sum(T(x+\delta+\alpha q)[pse'])}{k}$\;
    \If{$Pr_k'\leq Pr_k$}{
        $\delta = \delta + \alpha q$\;
        $Pr_k = Pr_k'$\;
        break\;
        }
    }
    }
    \Return $\delta$
\end{algorithm}

\subsubsection{Augmentation with adversarial examples}
\begin{figure*}[t]
\centering
\includegraphics[width=0.8\textwidth]{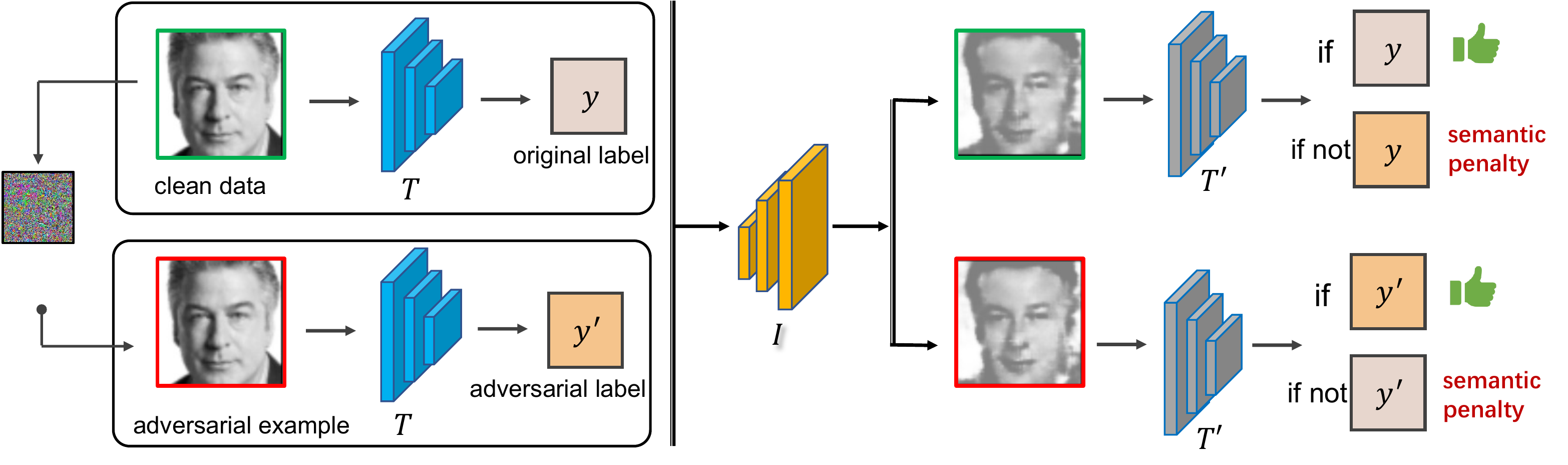} 
\caption{Explanation of the adversarial augmentation technique. Clean data with original labels and adversarial examples with adversarial labels are used to train the inversion model $I$. The semantic loss forces the inversion model to reconstruct diverse images for a clean image and an adversarial image. The reconstructed images should then be recognized as samples of the expected class. If not, the semantic loss penalizes the training process.
}
\label{fig:augment}
\end{figure*}

These produced perturbed examples are then fed into the target model, and the incorrectly predicted output, i.e., soft-labels, are recorded as adversarial labels. In this way, the augmented training data comprises both adversarial examples and adversarial labels. Fig.~\ref{fig:augment} explains the process. 
Given the black-box access to the target model, an adversarial image $x'$ is crafted by carefully adding noise to the image. $y'$ is the adversarial label of $x'$, which is the predicted soft-label when $x'$ is fed into the target model. 
Clean data $x$ with original labels $y$ and adversarial examples $x'$ with adversarial labels $y'$ are used to train the inversion model $I$. The semantic loss forces the inversion model to reconstruct diverse images for a clean image and an adversarial image. The reconstructed images should then be recognized as samples of the expected class. If not, the semantic loss penalizes the training process. 

The differences between $x$ and $x'$ mostly surround the class-related parts, as validated by Yang et al.~\cite{nips/YangZZTT21}. Both the clean and augmented adversarial data are used to train the inversion model. The class-redundant parts in these two types of data allow the reconstruction loss $\mathcal{L}_r$ to be minimized easily in the training phase. More importantly, the semantic loss $\mathcal{L}_s$ forces the training process to pay more attention to the class-related parts because these two types of data have similar class-redundant parts but different class-related parts. 

Without the regularization of $\mathcal{L}_s$, the adversarial augmentation might not improve attack performance. However, guided by our proposed loss function, the reconstructed version of clean data $x$ is forced to be classified into the original label $y$. Otherwise, the semantic loss $\mathcal{L}_s$ imposes a penalty. Noting that, for a machine classifier, the label of the adversarial example $x'$ should be the corresponding target class $y'$ instead of the original class $y$. Likewise, the reconstruction of the adversarial example $x'$ must be predicted as the corresponding adversarial label $y'$ instead of $y$ to avoid an excessive semantic loss $\mathcal{L}_s$. Intuitively, if a well-trained inversion model is able to reconstruct two similar images that a machine identifies as samples of different classes, we believe the inversion model has learned to capture the class-related parts that determine the classification tasks. Therefore, for the adversarial examples, the class-related information refers to the target class instead of the original class.

\subsection{Discussion}
In general, both the regularized loss function and the adversarial augmentation procedure aim to regularize training of the inversion model and prevent the model from being overfitted to the class-redundant information. Our regularized loss function explicitly adds a regularization term to prevent this from happening, while the adversarial augmentation procedure implicitly boosts the contribution of the class-related parts during training by increasing the variety of class-related parts.
\subsubsection{Adversarial Augmentation vs Adversarial Training}
The strategy of our adversarial augmentation is similar to that of adversarial training~\cite{iclr/MadryMSTV18}. However, the underlying ideas are completely different. As a defense against adversarial attacks, adversarial training schemes search for perturbations that will mislead the target model. The target model is then further trained on these perturbed data with \textbf{original labels}. The objective is to increase the robustness of the target model against this noise and decrease the likelihood of providing wrong predictions. Our adversarial augmentation technique adds perturbations to the class-related parts and assigns \textbf{adversarial labels} to these perturbed data, which boosts their variety by introducing adversarial labels. Our loss function forces the inversion model to be more sensitive to the class-related parts, resulting in better reconstructions of class-related parts and better reconstructed data that can be correctly classified by machine learning classifiers.
\subsubsection{More complex scenarios}
In our method, we applied a neural network with feature extractors related to the attack task to evaluate the quality of the reconstructed images, and the output of this neural network is used to calculate the semantic loss, which can further enhance the attack accuracy from a machine's perspective. Like the seminal work [1], we mainly focus on the setting where the target task is the same as the attack task. 
However, if the reconstructed samples are used for another task other than the task of the target model, this is a more complex attack scenario. In this case, we can modify the semantic loss. Specifically, we can use our neural network evaluator to assess the reconstruction quality for the attack task. For instance, if the reconstructed samples are used to fool identity classifiers while the target model is trained to classify emotion, we can compute the semantic loss using a classifier for identity. And then, the label $y$ should be the identity label. 
In addition, to enhance the reconstruction quality further, we can use a multi-view semantic loss involving both two tasks, i.e., the target task and the attack task. For instance, if the reconstructed samples are used to fool identity classifiers while the target model is trained to classify emotion, then the label $y_1$ should be the identity, and $y_2$ should be the emotion. The semantic loss can be formalized as follows:
$$L_s=\mathbb{E}_(x\in X) [{\rm CE}(T_1 (x),y_1 )+{\rm CE}(T_2 (x),y_2)]$$
where $T_1$ is the identity classifier and $T_2$ is the emotion classifier. 

\subsubsection{Pseudo-label's resistance to noises}
\begin{figure}[t]
\centering
\includegraphics[width=0.6\columnwidth]{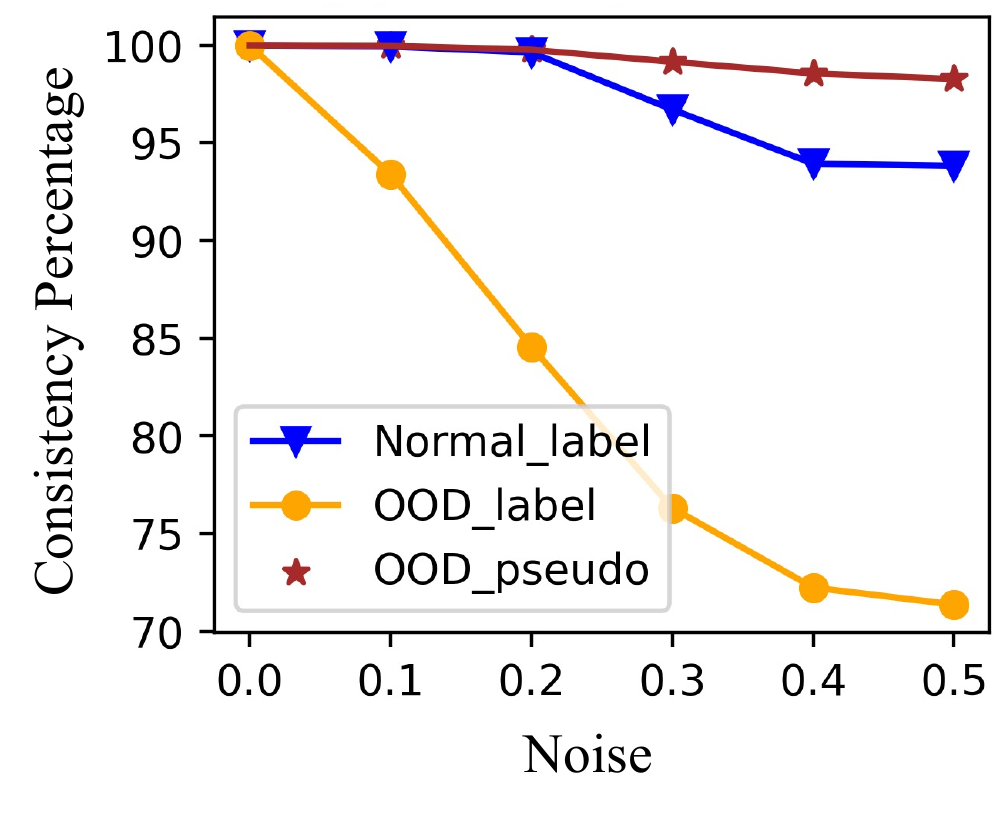} 
\caption{The percentage of randomly perturbed data that have the same predicted labels (i.e., top-$1$ label) or pseudo-labels as the clean data. The blue and orange lines, respectively, represent the impacts of random noises on the labels of normal data and out-of-distribution (OOD) data, where normal data have identical distribution as the training data of the classifier. The brown line represents the impacts of random noise on the pseudo-labels of the OOD data.}
\label{fig:consistency}
\end{figure}

In addition, we also empirically validated our pseudo-label's property of being resistant to random noises. More specifically, we calculated the percentage of randomly perturbed data that have the same predicted labels (i.e., top-$1$ labels) or pseudo-labels as the clean data in the inference phase on MNIST. First, we examined the impact of random noises on data with the same distribution as the classifier's training data. The findings in Fig.~\ref{fig:consistency} show that harmless random noises have little effect on the predicted labels. However, the out-of-distribution data reveal a distinct pattern. When the out-of-distribution data are perturbed with random noises, the top-$1$ labels are changed for a majority of data. As a result, the top-$1$ label is inappropriate to guide the generation of adversarial examples for out-of-distribution data. We also investigated how well our pseudo-labels kept against random noises on out-of-distribution data. The results demonstrate that our pseudo-labels on out-of-distribution data have similar characteristics as the top-$1$ labels on the identical-distribution data. The predicted pseudo-labels of out-of-distribution data are barely affected by non-malicious random noises. The larger $k$ is, the more resistant the pseudo-label is.

\section{Theoretical Analysis}
In this section, we explain our method from the theoretical standpoint. Our analysis comprises two parts: the ineffectiveness of the prior loss function and emphasis of the class-related parts of our method. The ineffectiveness of the previous loss function analysis shows why the MSE results in a poor reproduction of class-related parts in the training of the inversion model. And the emphasis of the class-related parts of our method demonstrates why our method can enhance the reproduction of class-related parts by regularizing the loss function and leveraging adversarial augmentation. 

\subsection{The ineffectiveness of MSE}
\label{sec:theo1}
Before the analysis, some assumptions from the previous work~\cite{nips/YangZZTT21} are presented.
\begin{assumption}
\label{assu:1}
A disentangling function $G$ exists such that an image $x$ can be divided to two parts by $G$: class-redundant part $xr=G(x)$ and class-related part $xc = x-G(x)$. In addition, ${\rm Pr}[T(x)=y]\approx{\rm Pr}[T(xc)=y]\gg{\rm Pr}[T(xr)=y]$. 
\end{assumption}

Yang et al.~\cite{nips/YangZZTT21} empirically proved that this function $G$ can be implemented using autoencoder models.
\begin{assumption}
\label{assu:2}
Let $xr$ and $xc$ be, respectively, the class-redundant part and class-related part, the pixels in $xr$ has a larger expectation and variance than pixels in $xc$. That is:
$\mathbb{E}(xr_k)>\mathbb{E}(xc_k)$ and $\mathbb{D}(xr_k)>\mathbb{D}(xc_k)$, where $xc_k$ and $xr_k$ are the $k$-th pixel of $xc$ and $xr$, respectively.
\end{assumption}

This assumption is supported by the experimental results of Yang et al.~\cite{nips/YangZZTT21}, which is also displayed in Fig.~\ref{fig:CDVAE}.
\begin{proposition}
If MSE is chosen as the metric in the training of the inversion model $I$, the reconstruction error of class-redundant part $xr$ of an image $x$ is less than class-related $xc$ in the reconstructed data. 
\end{proposition}
\begin{IEEEproof}
The output of $I$ is denoted as $I(T(x)) = \hat{x} = \hat{xr} + \hat{xc}$. We assumed that the original images' pixels $xr_k\sim \mathcal{N}(\mu_r, \sigma_r)$ and $xc_k\sim \mathcal{N}(\mu_c, \sigma_c)$. The training of the inversion model $I$ begins with a random initial state, and the output can also be thought of as two variables following the normal distributions. Specifically, $\hat{xr}\sim \mathcal{N}(\mu_r, \sigma_r)$ where $\mu_r = \mathbb{E}(xr)$ and $\sigma_r = \mathbb{D}(xr)$. Likewise, $\hat{xc}\sim \mathcal{N}(\mu_c, \sigma_c)$ where $\mu_c = \mathbb{E}(xc)$ and $\sigma_c = \mathbb{D}(xc)$. This is a valid statement because the inversion model can quickly learn the distribution of the original image, $xr$ and $xc$, after a few number of epochs. Typically, the expectation of MSE between the $k$-th pixel of $\hat{xr}$ and that of $xr$ can be stated as 
\begin{equation}
\begin{split}
	&\mathbb{E}_{\{xr_k, \hat{xr}_k \in \mathcal{N}(\mu_r, \sigma_r)\}}\left[{\rm MSE}(xr_k,\hat{xr}_k)\right] \\
	&\qquad= \mathbb{E}_{\{xr_k, \hat{xr}_k \in \mathcal{N}(\mu_r, \sigma_r)\}}[(xr_k-\hat{xr}_k)^2]\\
	&\qquad=\mathbb{E}_{\{xr_k, \hat{xr}_k \in \mathcal{N}(\mu_r, \sigma_r)\}}^2\left[xr_k-\hat{xr}_k\right] \\
	&~~\qquad+ \mathbb{D}_{\{xr_k, \hat{xr}_k \in \mathcal{N}(\mu_r, \sigma_r)\}}(xr_k-\hat{xr}_k)\\
	&\qquad= 0 + 2\sigma_r^2\\
	&\qquad=2\sigma_r^2.
\end{split}
\end{equation}
Similarly, the expectation of MSE between the $k$-th pixel of $\hat{xc}$ and that of $xc$ is $2\sigma_c^2$. 

We assume that the inversion model $I$ at $N$-th epoch can accurately reconstruct $n$ pixels, with $n_1$ pixels distributed to $\hat{xr}$ and $n_2 = n-n_1$ pixels allocated to $\hat{xc}$. If $n$ is fixed, there is a trade-off between $n_1$ and $n_2$. In this case, the MSE between the original image $x$ and its reconstructed version $\hat{x}$ can be calculated as follows:
\begin{equation}
\label{equ:mse}
\begin{split}
	&{\rm MSE}(x,\hat{x}) = \sum\nolimits_k^K(x_k-\hat{x}_k)^2 \\
	&= \sum\nolimits_k^K(xr_k+xc_k-\hat{xr}_k - \hat{xc}_k)^2 \\
	&= \sum\nolimits_k^K((xr_k-\hat{xr}_k)^2 + (xc_k-\hat{xc}_k)^2)\\ 
	&\qquad\qquad+ \sum\nolimits_k^K(2(xr_k-\hat{xr}_k)(xc_k-\hat{xc}_k))\\
	&= \sum\nolimits_k^K({\rm MSE}(xr_k,\hat{xr}_k)+{\rm MSE}(xc_k,\hat{xc}_k))\\
	&\qquad\qquad+ \sum\nolimits_k^K(2(xr_k-\hat{xr}_k)(xc_k-\hat{xc}_k))\\
	&= n_1\cdot 0+(K-n_1)\cdot 2\sigma_r^2 +n_2\cdot 0+(K-n_2)\cdot 2\sigma_c^2\\
	&\qquad\qquad+ \sum\nolimits_k^K(2(xr_k-\hat{xr}_k)(xc_k-\hat{xc}_k))\\
\end{split}
\end{equation}
For the $n_1$ pixels that are accurately reconstructed in $\hat{xr}$, the last term in Equation~\ref{equ:mse}, $2(xr_k-\hat{xr}_k)(xc_k-\hat{xc}_k)$, equals $0$ because the $xr_k=\hat{xr}_k$. Likewise, for the $n_2$ pixels that are accurately reconstructed in $\hat{xc}$, this term also equals to $0$. Consequently, the value of this term is not impacted by the trade-off between $n_1$ and $n_2$, and can be represented by $C$. The MSE in Equation~\ref{equ:mse} can be approximated as follows:
\begin{equation}
\begin{split}
	&{\rm MSE}(x,\hat{x}) \\
	&= (K-n_1)\cdot 2\sigma_r^2 +(K-n_2)\cdot 2\sigma_c^2 + C\\
	&= 2K(\sigma_r^2+\sigma_c^2) - 2n_1\cdot\sigma_r^2 - 2n_2\cdot\sigma_c^2 + C\\
	&= 2K(\sigma_r^2+\sigma_c^2) - 2n_1\cdot\sigma_r^2 - 2(n-n_1)\cdot\sigma_c^2 + C\\
	&= 2K(\sigma_r^2+\sigma_c^2) - 2n\cdot\sigma_c^2 - 2n_1\cdot(\sigma_r^2 -\sigma_c^2) + C\\
	&= 2K(\sigma_r^2+\sigma_c^2) - 2n\cdot\sigma_r^2 - 2n_2\cdot(\sigma_c^2 -\sigma_r^2) + C\\
\end{split}
\end{equation}
Given the variance of class-redundant part of an image is larger, i.e., $\sigma_r>\sigma_c$ (Assumption~\ref{assu:2}), the MSE on training data is negative with $n_1$ and positive with $n_2$. Consequently, when MSE is selected as the metric to guide the training of the inversion model, the training process forces $I$ to achieve a larger $n_1$ and a smaller $n_2$. Namely, more pixels that are accurately reconstructed are allocated in the class-redundant part $xr$ than $xc$. As a result, the reconstruction error of $xr$ is less than $xc$ when the inversion model is trained with MSE loss.
\end{IEEEproof} 

The greater reconstruction error of the class-related parts $xc$ leads to worse semantic information in the reconstructed data, therefore, the attack accuracy of the reconstructed data is not high when inversion model is trained using MSE loss. 

\subsection{The emphasis of class-related parts of our method}

We replace the MSE with a regularized loss function to address the imbalance between the reconstructions of class-related and class-redundant parts. And then, our adversarial augmentation is used to further improve the reproduction of the class-related parts (i.e., semantic information).
\begin{proposition}
When compared to the unregularized loss function $\mathcal{L}_r$, our regularized loss function $\mathcal{L} = \mathcal{L}_r + \lambda\cdot\mathcal{L}_s$ can decrease the reconstruction error of class-related parts $xc$ than class-redundant part $xr$, and hence, enhance the reconstruction quality of class-related parts of the original data $x$.
\end{proposition}
\begin{IEEEproof}
Given $xr = G(x)$ and $xc=x-G(x)$, the class-related part $xc$ is more predictive than the class-redundant part $xr$ as shown in Assumption~\ref{assu:1}, that is, ${\rm Pr}[T(xc)=y]\gg{\rm Pr}[T(xr)=y]$. For the training data $(X,Y) = (XR, XC, Y)$, we denote the conditional entropy as follows: $$H(Y|XC)=\sum_{y\in Y}\sum_{xc\in XC}(Pr(y_i|xc)),$$
$$H(Y|XR)=\sum_{y\in Y}\sum_{xr\in XR}(Pr(y_i|xr)).$$ 
As a result, $H(Y|XR) > H(Y|XC)$. The mutual information between the class-related part $xc$ and the label $y$ can be calculated: $$MU(XC,Y) = H(Y)-H(Y|XC).$$ Because the conditional entropy between $XC$ and label $Y$ is larger, $MU(XC,Y)>MU(XR,Y)$. Consequently, for a well-trained classifier $T$ and the disentangling function $G$, the cross entropy ${\rm CE}(T(xc),y)$ should be extremely minor while ${\rm CE}(T(xr),y)$ might be not small. Therefore, when we try to minimize the cross entropy of the prediction of reconstructed data and the ground truth label (i.e., ${\rm CE}(T(\hat{x}),y)$), the cross entropy ${\rm CE}(T(\hat{xc}),y)$ can converge quickly to a low level by optimizing the reconstruction of $\hat{xc}$ when compared to ${\rm CE}(T(\hat{xr}),y)$. Our regularized loss function can be formulated as follows:
\begin{equation}
\begin{split}
    \mathcal{L} &= \mathcal{L}_r + \lambda\cdot\mathcal{L}_s\\
    &=~{\rm MSE}(\hat{x}, x) + \lambda\cdot {\rm CE}(T(\hat{x}),y)\\
    &\Leftrightarrow~{\rm MSE}(\hat{x}, x) + \lambda\cdot {\rm CE}(T(\hat{xc}),y).\\
\end{split}
\end{equation}
If $T$ is well-trained, the minimization of ${\rm CE}(T(x),y)$ can be thought as the minimization of $ {\rm CE}(T(xc),y)$. And the minimization of $\mathcal{L}_s$ forces $T(\hat{xc})$ to approach $T(xc)$. 

If the inversion model $I$ at $N$-th epoch can accurately reconstruct $n$ pixels, with $n_1$ pixels distributed to $xr$ and $n_2 = n-n_1$ pixels allocated to $xc$, then a larger $n_2$ implies a greater degree of resemblance between $xc$ and $\hat{xc}$. Consequently, $\mathcal{L}_s\Leftrightarrow~\lambda\cdot {\rm CE}(T(\hat{xc}),y)$ is inversely proportional to $n_2$, and minimizing $\mathcal{L}_s$ leads to a smaller $n_1$ and a larger $n_2$, which compensates for the imbalance between the reconstruction of $xr$ and $xc$ caused by MSE. The reconstruction error of class-related parts $xc$ can be decreased further, and the likelihood that reconstructed data are correctly classified is subsequently increased.
\end{IEEEproof}
\begin{assumption}
\label{assu:3}
Let $x'$ be the adversarial example of $x$, then the adversarial perturbation $x'-x$ is bounded by a small constant $\delta$. In addition, $x'-x\approx xc'- xc$, and $xr'\approx xr$, where $xc'$ and $xr'$ represent the class-related and class-redundant part of $x'$, respectively.
\end{assumption}
Yang et al.~\cite{nips/YangZZTT21} empirically validated the adversarial perturbations are predominantly allocated to the class-related part while constructing adversarial examples, which is the support of Assumption~\ref{assu:3}. Inspired by the study on quantifying mechanisms of data augmentation~\cite{DataAug2020}, we offer a similar metric, Diversity, to assess the performance of data augmentation in data reconstruction tasks. Higher diversity of augmented data leads to better generalization of the model trained on these augmented data.
\begin{definition}[Diversity]
Let $\mathcal{A}$ be an augmentation and $D_{trn}'$ be the augmented training data resulting from applying the augmentation $\mathcal{A}$ to the original training data $D_{trn}$. Further, let $L_{trn}'$ be the training loss for a model, $m$, trained on $D_{trn}'$, and $L_{trn}'$ be the training loss trained on $D_{trn}'$. We define the Diversity, $Div[D_{trn}; D_{trn}']$, as
\begin{equation}
    \begin{split}
    \nonumber
        Div[D_{trn}, D_{trn}']=&\mathbb{E}_{x\in D_{trn}'}\left[L_{trn}'(x, \hat{x})\right] \\
        & - \mathbb{E}_{x\in D_{trn}}\left[L_{trn}(x, \hat{x})\right].
    \end{split}
\end{equation}

\end{definition}

\begin{proposition}
Adversarial augmentation results in higher diversity in terms of $\mathcal{L}_s$ than $\mathcal{L}_r$.
\end{proposition}
\begin{IEEEproof}
Our training loss $\mathcal{L}$ consists of two components: the reproduction loss $\mathcal{L}_r$ and the semantic loss $\mathcal{L}_s$. The augmented data $X'$ results in different degrees of improvement for these two losses. In particular,  
\begin{equation}
\label{div}
    \begin{split}
        Div&[X, X']\\
        &=\mathbb{E}_{x\in X'}\left[L_{trn}'(x, \hat{x})\right]-\mathbb{E}_{x\in X}\left[L_{trn}(x, \hat{x})\right]\\
        &=\mathbb{E}_{x\in X'}\left[\mathcal{L}'_r+\lambda\cdot L\mathcal{L}'_s\right]\\
        &~\qquad - \mathbb{E}_{x\in X}\left[\mathcal{L}_r+\lambda\cdot L\mathcal{L}_s\right]\\
        &=\mathbb{E}_{x\in X'}\left[\mathcal{L}'_r\right] - \mathbb{E}_{x\in X}\left[\mathcal{L}_r\right]\\
        &~\qquad + \mathbb{E}_{x\in X'}\left[\lambda\cdot \mathcal{L}'_s\right] - \mathbb{E}_{x\in X}\left[\lambda\cdot \mathcal{L}_s\right]\\
    \end{split}
\end{equation}
Revisiting our adversarial augmentation, the new augmented data $x'$ (i.e., the adversarial example) is visually indistinguishable to the original data $x$, and thus, the training loss in terms of the MSE on the augmented data is not affected significantly. As a result, the diversity for the reconstruction loss $\mathcal{L}_r$ in the training process $\mathbb{E}_{x\in X'}\left[\mathcal{L}'_r\right] - \mathbb{E}_{x\in X}\left[\mathcal{L}_r\right]\approx 0$. However, there is a different case for the semantic loss $\mathcal{L}_s$. 
    
As shown in Assumption~\ref{assu:3}, our adversarial augmentation provides more augmentation on $xc$ than $xr$. Though the perturbations $xc'-xc$ is still small, the corresponding adversarial label $y'=\arg\max_i T_i(xc')$ is completely different from the original label $y=\arg\max_i T_i(xc)$. Consequently, the training loss in terms of the $\mathcal{L}_s$ on the augmented data might be increased greatly due to the disparity between the perturbations of data (i.e., minor $xc'-xc$) and the changes of predicted labels (i.e., different $y'$ and $y$). In other words, $\mathbb{E}_{x\in X'}\left[\cdot \mathcal{L}'_s\right] - \mathbb{E}_{x\in X}\left[\cdot \mathcal{L}_s\right]$ is large. Hence, the diversity of our adversarial augmentation is dominated by the semantic loss $\mathcal{L}_s$ rather than the reconstruction loss $\mathcal{L}_r$. As a result, the augmented training data resulting from applying our adversarial augmentation effectively increases the diversity for the semantic loss $\mathcal{L}_s$ compared to the reconstruction loss $\mathcal{L}_r$, which provides more knowledge for optimizing the semantic loss $\mathcal{L}_s$ and forces the inversion model to pay more attention to the semantic loss.
\end{IEEEproof}

\section{Performance Evaluation}
We conducted extensive experiments on two datasets and two neural network architectures. We assumed that the attacker does not know the target classifier's architecture or parameters and does not have access to the same class of data as the private training data. Namely, there is no overlap between the private data and the attack data.

\subsection{Datasets and Experimental Setup}
\subsubsection{Datasets}
Following~Yang et al.'s work~\cite{ccs/YangZCL19}, we experimented with two datasets, MNIST~\cite{lecun-mnist} and FaceScrub~\cite{icip/facescrub}. Each dataset was divided into two parts: the target data (i.e., private data), used to train the target model, and the attack data (i.e., public data), used to train the inversion model. To this end, we selected photos of $100$ people from FaceScrub as the private data, which we used to train the target model, and photos of another $100$ people as the public data, which we used to train the shadow and the inversion models. With the MNIST dataset, we divided the dataset into two parts, the digits $0\sim4$ and the digits $5\sim9$. The former was used as private data, and the latter was used as public data.   
\subsubsection{Baseline}
As the seminal learning-based model inversion attack, Yang et al.'s attack~\cite{ccs/YangZCL19} is commonly used as a benchmark for model inversion performance. In this paper, as with the seminal work~\cite{ccs/YangZCL19}, we focus on scenarios where the target task and attack task are the same, with no auxiliary tasks present. In these cases, only the outputs of the target model can be used for reconstruction. Zhao et al.'s attack~\cite{iccv/ZhaoZXL21}, which introduces model explanations, focuses on a different situation to ours. Specifically, the target task is different from the attack task, and the reconstruction process requires additional auxiliary information, such as explanations of target models. It should be noted that the attacker is assumed to have access to model explanations of the target model, which may not be practical in some situations. Therefore, Zhao et al.'s work~\cite{iccv/ZhaoZXL21} is not included as a baseline. 
\subsubsection{Evaluation Metrics}
We relied on two widely-used metrics to evaluate the efficacy of our model inversion attack: reconstruction error and attack accuracy~\cite{ccs/YangZCL19,kahla2022label}. The former measures the pixel-wise differences between the original and the reconstructed images, and lower reconstruction error indicates higher reconstruction quality and attack effectiveness. The latter refers to the ratio of reconstructed images that are correctly classified by machine learning models. Greater attack accuracy equates to better attack performance and a greater potential for downstream attacks, which is the primary objective of this paper. 
\subsubsection{Architecture}
The architectures used were a $3$-layer and a $4$-layer CNN~\cite{ccs/YangZCL19} and ResNet-$18$~\cite{he2016identity}, as the structure of the target models. For a fair comparison, the architecture of our inversion model was identical to that of the baseline~\cite{ccs/YangZCL19}. In addition, to avoid the reconstructed data over-fitting to the shadow models, the structure of the evaluation models should be different from that of the shadow model. Therefore, we train a VGG-$11$~\cite{vgg} based on the private training data as an evaluation model to assess the attack accuracy.

\subsection{Experimental Results}
\subsubsection{Ablation Study}
\begin{figure}[t]
\centering
\includegraphics[width=0.98\columnwidth]{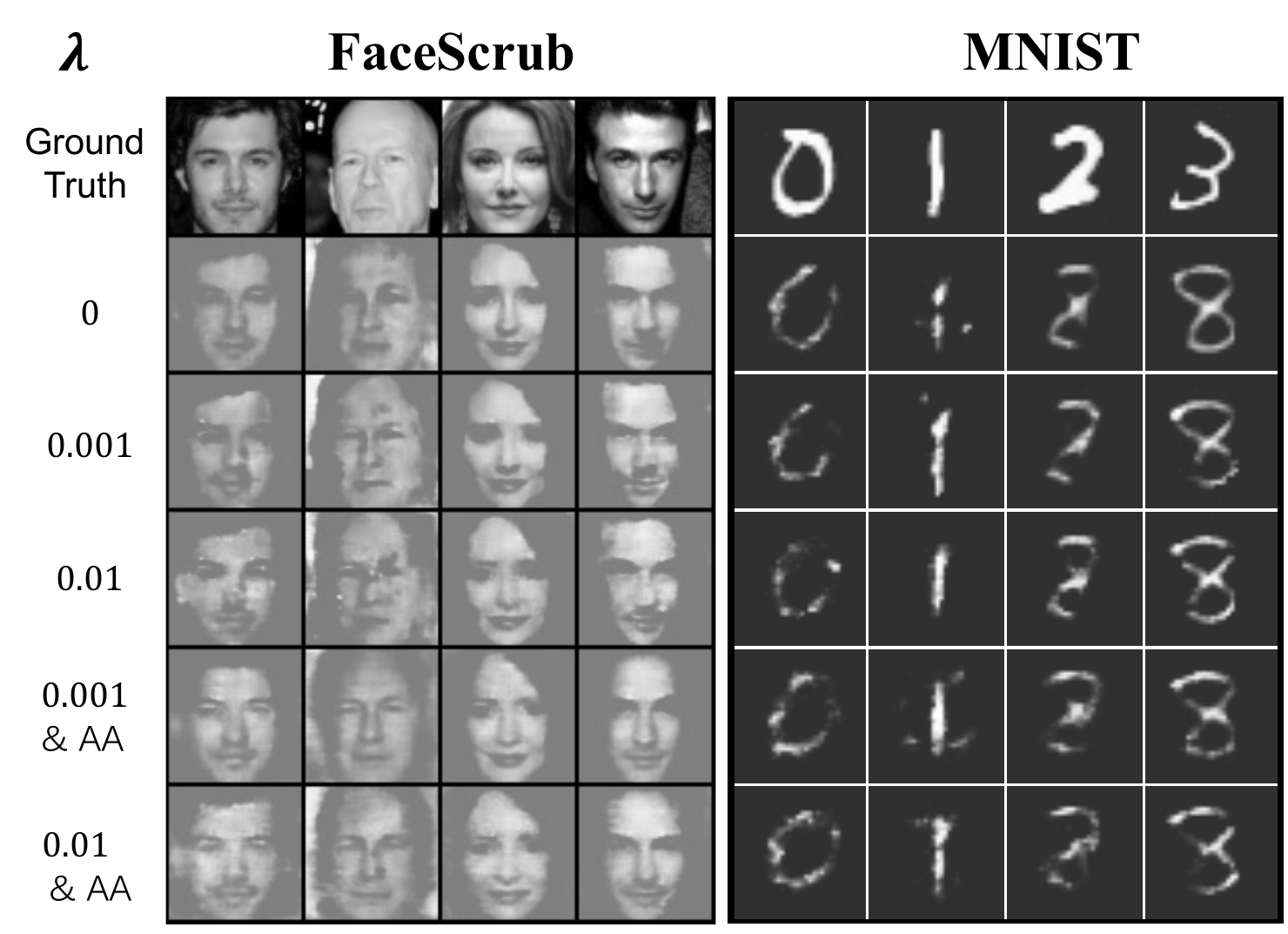} 
\caption{Comparison of the reconstructed images. $\lambda=0$ represents the baseline. The top two rows show the results with the regularized loss function and the adversarial augmentation, while the following two rows demonstrate the results using only the regularized loss function.}
\label{fig:lambda}
\end{figure}

Our method consists of two techniques: the additional semantic loss $\mathcal{L}_s$ and the adversarial augmentation. To validate their effectiveness, we conducted ablation experiments using FaceScrub and MNIST in the worst-case setting, where the target model, the shadow model, and the evaluation model all have completely different architectures. Specifically, we use the architectures designed by Yang et al.~\cite{ccs/YangZCL19} as the target models. A ResNet-$18$ is trained as the shadow model, and VGG-$11$ is selected as the structure of the evaluation model. 
\begin{table*}[t]
\renewcommand{\arraystretch}{1.3}
\centering
\caption{Performance with varying values for $\lambda$, which is the coefficient of the semantic loss $\mathcal{L}_s$. $\lambda=0$ is the baseline, which means the semantic loss and adversarial augmentation techniques are not used. `Error' denotes the reconstruction error, and `Acc.' denotes the attack accuracy.}
\label{tab:lambda}
\begin{tabular}{l|ll|ll|ll||ll|ll|ll}
	\hline
	\textbf{Setting}&\multicolumn{4}{l}{Target Model: $3$-layer/$4$-layer CNN}&\multicolumn{4}{c}{Shadow Model: ResNet-$18$}&\multicolumn{4}{r}{Evaluation Model: VGG-$11$}\\
	\hline
	\textbf{Dataset} &\multicolumn{6}{c||}{FaceScrub}&\multicolumn{6}{c}{MNIST}\\
	\hline
    \textbf{$\lambda$}  & \multicolumn{2}{c|}{$0.01$} & \multicolumn{2}{c|}{$0.005$}&\multicolumn{2}{c||}{$0.001$}& \multicolumn{2}{c|}{$0.01$} & \multicolumn{2}{c|}{$0.005$}&\multicolumn{2}{c}{$0.001$}\\
    & Error&Acc.& Error&Acc.& Error&Acc.& Error&Acc.& Error&Acc.& Errors&Acc. \\ 
    \hline
    $\mathcal{L}_r$~($\lambda=0$)& \multicolumn{2}{c}{}&$\textbf{0.203}$& \multicolumn{1}{c}{$12.4\%$}& \multicolumn{2}{c||}{}& \multicolumn{2}{c}{}&$\textbf{0.882}$& \multicolumn{1}{c}{$72.5\%$}&\multicolumn{2}{c}{}\\
    \hline
    $\mathcal{L}_r$~\&~$\mathcal{L}_s$&$0.211$&$20.4\%$&$0.209$&$20.2\%$&$0.207$&$19.6\%$&$0.885$&$90.4\%$&$0.882$&$89.2\%$&$0.883$&$76.1\%$\\
    $\mathcal{L}_r$~\&~$\mathcal{L}_s$~\&~AA&$0.240$&$\textbf{31.7}\%$&$0.227$&$30.4\%$&$0.217$&$25.4\%$&$0.887$&$\textbf{94.1}\%$&$0.887$&$94.0\%$&$0.890$&$87.9\%$\\
    \hline
\end{tabular}
\end{table*}
We trained multiple inversion models using $\mathcal{L}_r$, $\mathcal{L}_r$\&$\mathcal{L}_s$, and $\mathcal{L}_r$\&$\mathcal{L}_s$\&AA (\textbf{A}dversarial \textbf{A}ugmentation). It is worth noting that the effectiveness of the AA is based on the constraint of $\mathcal{L}_s$. AA without $\mathcal{L}_s$ provides no positive effects, and the results of $\mathcal{L}_r$\&AA will be omitted. We compared the reconstructed images, reporting the reconstruction errors in terms of MSE and attack accuracy, as shown in Fig.~\ref{fig:lambda} and Table~\ref{tab:lambda}. Here, $\lambda$ controls the tradeoff between reconstruction loss and the semantic loss, with a larger $\lambda$ allowing the inversion model to pay more attention to the semantic reproduction instead of the pixel-wise reconstruction error. As shown in Table~\ref{tab:lambda}, without adversarial augmentation, the additional $\mathcal{L}_s$ still increased the attack accuracy of the reconstructed data with a minor improvement of the reconstruction error. Additionally, the adversarial augmentation further improved the attack accuracy while only slightly increasing the reconstruction error. However, as demonstrated in Fig.~\ref{fig:lambda}, this slight increase in the reconstruction error does not significantly impact the visual quality of the reconstructed images.

In addition, the visual results in Fig.~\ref{fig:lambda} demonstrated that our technique with the added semantic loss and AA could capture more class-related parts in reconstruction tasks. Specifically, Yang et al's technique (i.e., the baseline)~\cite{ccs/YangZCL19} trained an inversion model based on the public attack data, resulting in images that are more likely to mimic the attack data distribution than the private distribution. Given a confidence vector for the image of the digit $3$, for instance, the reconstructed data produced by the baseline inversion model is more likely to be recognized as the digit $8$. In comparison, our technique with additional semantic loss and AA can reconstruct an image resembling the digit $3$ despite the fact that our inversion model has never seen photos of the digit $3$. Because of this, our technique can achieve higher attack accuracy.

\subsubsection{Class Activation Map}
\begin{figure*}[t]
\centering
\includegraphics[width=0.9\textwidth]{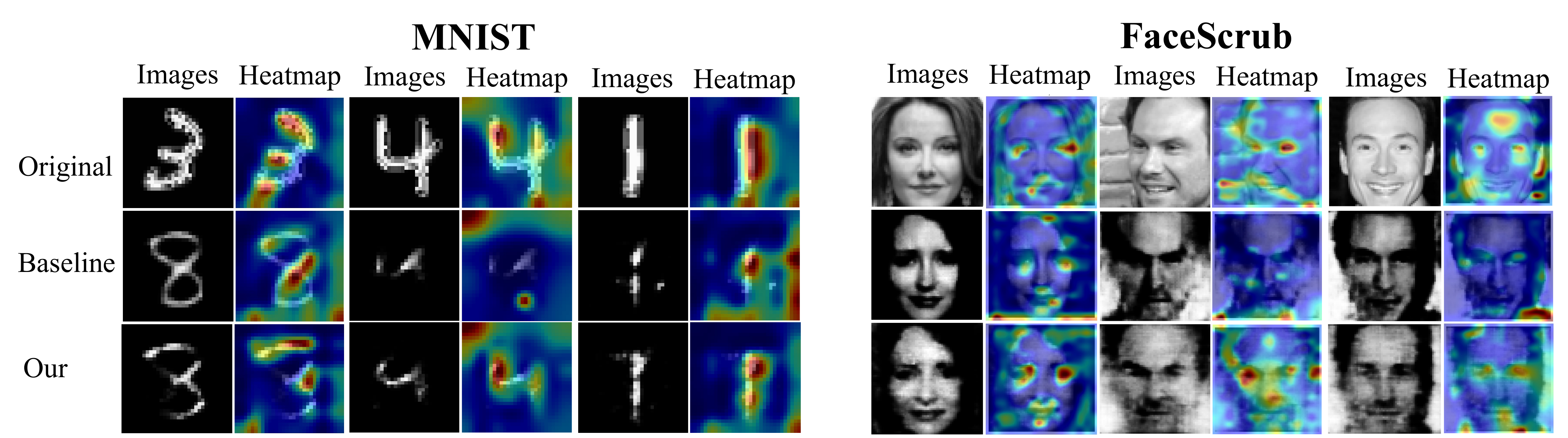} 
\caption{Comparison of the class activation map (CAM)~\cite{cam,torcham2020} of the reconstructed images. The first row is the private images and the corresponding heatmap. The second row is the reconstructed images using the baseline method and the corresponding heatmap. The third row is the reconstructed images using our method and the corresponding heatmap. The CAM of our method's reconstructed data are more likely to match that of the private data when compared to the reconstructed data of the baseline.}
\label{fig:cam}
\end{figure*}

To demonstrate the efficacy of our method in capturing class-related parts in images, we also compared the class activation map (CAM)~\cite{cam} of different reconstructed data, which identifies the crucial region-related classification tasks in an image. We employed the torch-cam library~\cite{torcham2020} to generate the heatmap and applied it to MNIST and FaceScrub. The results are illustrated in Fig.~\ref{fig:cam}. 

Compared to the complicated face photos in FaceScrub, the classifier can easily locate the significant features for classifying the images in MNIST due to the simplicity of the images. The CAM of the original private image in MNIST is identifiable and well-clustered. Nevertheless, the CAM of the reconstructed data of the baseline approach~\cite{ccs/YangZCL19} indicates that the classifier is confused while classifying this reconstructed image, and that some attention is given to the image's edge. In contrast, our reconstructed data has an aggregated CAM, and the confusing feature region (e.g., the clearly different region of the digits $3$ and $8$) is not covered by the CAM, similar to the original image. The FaceScrub results also indicate that the CAM of our reconstructed data is more likely to match the CAM of the private data.
\subsubsection{Performance on Different Models}
We also evaluated our methods on two target models with different architectures - a $4$-layer CNN ($4$-CNNs) and a ResNet-18 (ResNet) - and with three different evaluation models. This experiment which was performed on the FaceScrub dataset was designed to validate the generalizability of the attacks on different architectures. Two scenarios are considered: the worst-case and the case where the architecture is known. In the worst-case setting, the target model has a different architecture to the shadow model because the attacker only has black-box access. With a $4$-layer CNN as the target model, we used a ResNet shadow model, and vice versa. In the known architecture scenario, the attacker obviously trains a shadow model with the same architecture as the target model.

From Table~\ref{tab:models}, it is clear that our method greatly improves attack accuracy compared to that of the baseline regardless of the architectures in play. And the degree of performance enhancement is largely dependent upon the evaluation model. In particular, with the $4$-layer CNN as the target model, the attack yields an accuracy of over $50\%$ when the shadow model and the evaluation model share the same architecture. Even when the target model, shadow model, and evaluation model have completely different structures, the increase in attack accuracy is still greater than $1.5$ times. 

With ResNet as the target model, reconstruction tasks are more difficult than with $4$-CNNs as the target model, because ResNet has more complex feature extractors. Thus, the reconstructed data are more difficult to be recognized. In addition, the worst-case attack accuracy of reconstructed data on evaluation model VGG is greater than the known architecture scenario, as the shadow model with a different structure improves the generalization of the inversion model across different evaluation models. Moreover, given an evaluation model, the attack achieves the highest level of attack accuracy when our shadow model and the evaluation model share the same architecture.
\begin{table}
    \renewcommand{\arraystretch}{1.3}
    \caption{Attack performance comparison on different target models and evaluation models with various architectures on the FaceScrub dataset. `Ours\_wc' and `ours\_ka' represent the results in the worst-case and known target architecture scenarios, respectively. }
    \label{tab:models}
    \centering
    \begin{tabular}{ll|cccc}
         \hline
        Target&\multirow{2}{*}{Method}&\multirow{2}{*}{Error}&\multicolumn{3}{c}{Evaluation Model}\\
          Model&&&4-CNNs&ResNet&VGG\\
         \hline
         \multirow{2}{*}{4-CNNs}&baseline&$0.203$&$7.4\%$&$25.4\%$&$12.4\%$ \\
         &ours\_wc& $0.240$&$29.5\%$ &$\textbf{55.2}\%$ &$\textbf{31.8}\%$  \\
         &ours\_ka& $0.230$&$\textbf{65.7}\%$ &$51.9\%$ &$\textbf{31.8}\%$  \\
         \hline
         \multirow{2}{*}{ResNet}&baseline&$0.228$&$5.8\%$&$16.9\%$&$7.2\%$\\
          &ours\_wc&$0.238$& $\textbf{16.0}\%$&$57.3\%$&$\textbf{24.3}\%$  \\
          &ours\_ka& $0.232$&$12.3\%$ &$\textbf{75.7}\%$ &$15.2\%$  \\
         \hline
    \end{tabular}
\end{table}

\subsubsection{Effect of Adversarial Examples}
Since our method partly benefits from using adversarial examples via data augmentation, we conducted an experiment to show how the adversarial examples affect the attack's performance. Here, we compared how effective the adversarial examples were at fooling the target model, along with the attack performance of the inversion model (which had been trained on those examples). As shown in Table~\ref{tbl:adv}, the quality of the adversarial examples is crucial to the training of our inversion model. As the maximal number of rounds of the simple black-box adversarial attack (Max\_Round) gradually increases, the perturbation size (Average $l_2$) also increases, along with the increased attack success rate (ASR) of the generated adversarial examples. A similar trend can also be observed for the size of added noise in each iteration $\epsilon$. Hence, attack accuracy is strongly influenced by the adversarial samples, but the relationship between the ASR and the attack accuracy is still unclear. Therefore, the parameters of the generated adversarial samples need to be adjusted manually to allow the inversion model to achieve the best performance.

\begin{table}
    \renewcommand{\arraystretch}{1.3}
    \caption{Attack performance based on different augmented adversarial examples.  }
    \label{tbl:adv}
    \centering
    \begin{tabular}{c|l|cccc}
         \hline
         Max\_Round &$\epsilon$& Average $l_2$ & ASR & Error & Acc.\\
         \hline
         $100$ &\multirow{4}{*}{$0.2$}& $1.14$&$79.9\%$ & $0.235$& $32.9\%$\\
         $200$ & & $2.17$&$81.7\%$ & $0.235$& $\textbf{34.9\%}$\\
         $500 $ & & $5.38$&$89.8\%$ & $0.229$&$33.0\%$ \\
         $1000$ & & $9.71$&$\textbf{96.6}\%$ & $0.231$&$19.9\%$ \\
         \hline
         \multirow{4}{*}{$200$}& $0.05$  & $0.60$&$79.3\%$ & $0.230$&$29.9\%$\\
          & $0.1$ &$1.14$ & $80.2\%$ & $\textbf{0.223}$ & $30.1\%$\\
          & $0.15$ &$1.66$ & $81.1\%$ & $0.233$ & $24.7\%$\\
          & $0.25$ &$2.70$ & $82.4\%$ & $0.230$ & $25.7\%$\\
         \hline
    \end{tabular}
\end{table}

\subsubsection{Less Query Budget}
In addition, the generation of adversarial examples needs a high volume of queries to the black-box target model. In practice, queries to the online model are usually expensive, and some online APIs (e.g., Google's cloud vision API) limit the number of queries per minute. And the excessive volume of queries will increase the probability of being detected by the defense system. Therefore, we investigate the feasibility of querying the white-box shadow model instead of the black-box target model in Table~\ref{tbl:adv_shadow}. As we can see, when the maximal number of rounds is less than $500$, the attack success rate tends to be lower than querying the target model directly. Likewise, the final perturbation produced by querying the shadow model has a norm that is twice as large as that of the perturbation produced by querying the target model. In other words, directly querying the target model can craft adversarial examples with smaller perturbations, which achieve even greater attack success rates. However, we are still able to generate some effective adversarial examples against the target model by only querying the shadow model - at least, enough to train an inversion model with a high attack accuracy. By querying the shadow model, the query cost will decrease significantly at the expense of a negligible reduction of attack accuracy, making this attack highly practical in the real world.

\begin{table}
    \renewcommand{\arraystretch}{1.3}
    \caption{Attack performance based on adversarial examples crafted by querying the shadow model.}
    \label{tbl:adv_shadow}
    \centering
    \begin{tabular}{c|l|cccc}
         \hline
         Max\_Round &$\epsilon$& Average $l_2$ & ASR & Error & Acc.\\
         \hline
         $100$ &\multirow{4}{*}{$0.2$}& $2.23$&$69.2\%$ & $0.242$& $31.5\%$\\
         $200$ & & $4.30$&$71.6\%$ & $0.240$& $\textbf{33.3\%}$\\
         $500 $ & & $10.98$&$85.1\%$ & $0.238$&$26.9\%$ \\
         $1000$ & & $20.18$&$\textbf{97.5}\%$ & $0.240$&$28.6\%$ \\
         \hline
         \multirow{4}{*}{$200$}& $0.05$  & $1.10$&$67.7\%$ & $\textbf{0.232}$&$27.4\%$\\
          & $0.1$ &$2.18$ & $69.3\%$ & $0.234$ & $27.8\%$\\
          & $0.15$ &$3.24$ & $70.5\%$ & $0.241$ & $26.3\%$\\
          & $0.25$ &$5.38$ & $72.5\%$ & $0.241$ & $28.3\%$\\

         \hline
    \end{tabular}
\end{table}
\subsubsection{Attack Adversarially-trained Models}
\begin{figure}
    \centering
    \includegraphics[width=0.8\columnwidth]{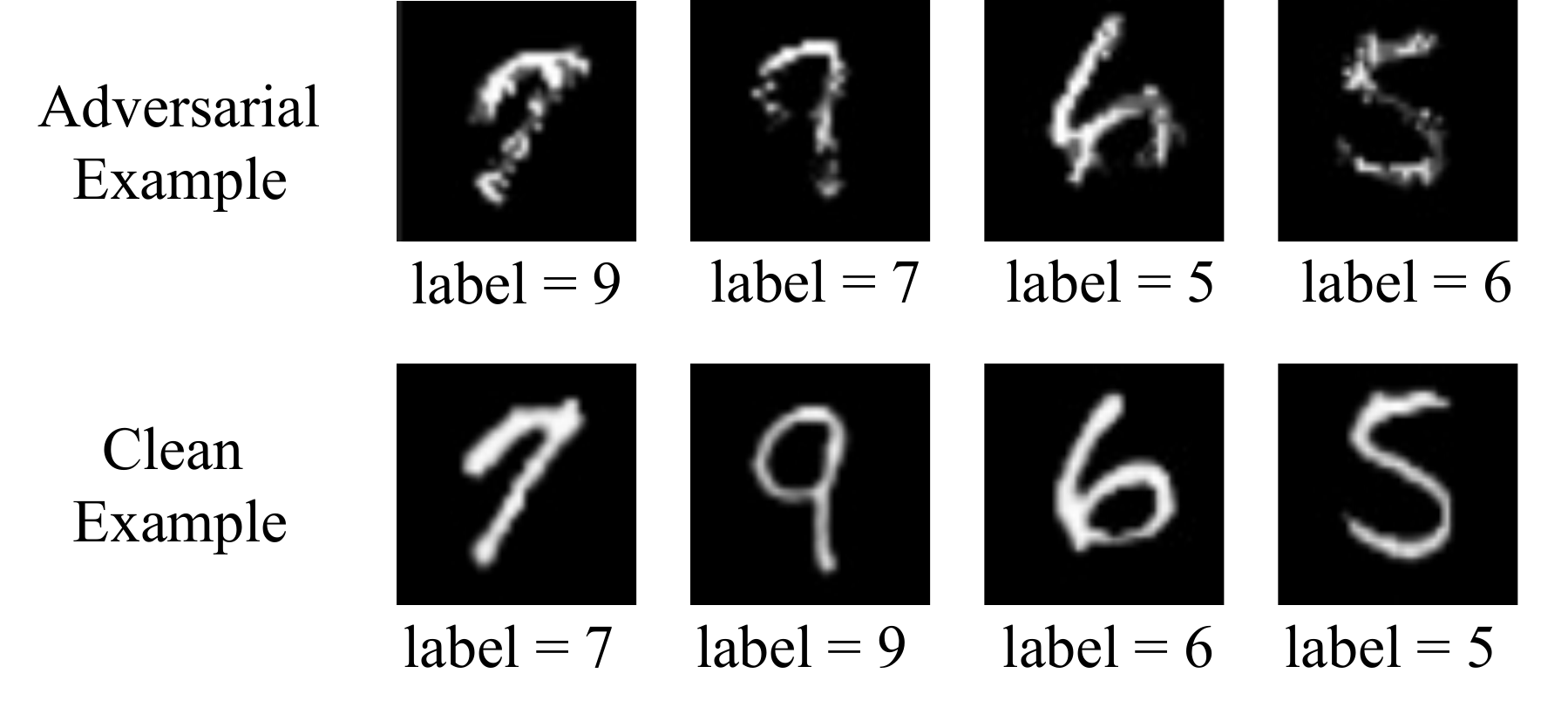}
    \caption{Some successful adversarial examples against an adversarially-trained model. The first row is the adversarial examples and the second row is the corresponding clean images.}
    \label{fig:adv_train_img}
\end{figure}

When the victim model is adversarially-trained, generating effective adversarial samples for these adversarially-trained models is indeed challenging in typical adversarial attacks. This is because the perturbation size is always restricted to ensure that the perturbed image is indistinguishable from the original image to the human perception. The perturbation required for adversarially-trained models is typically large and often exceeds the restriction of perturbation size, making it difficult to deceive the classifier successfully with imperceptible perturbations. However, in our scenario, we do not need the perturbed image to be visually indistinguishable from the original image. Therefore, even if the perturbation slightly exceeds the restriction of perturbation size, it is also acceptable in our method. Thus, we can easily obtain effective adversarial samples by increasing the size of the perturbation in this situation. We presented some successful adversarial examples against a model adversarially-trained on MNIST dataset in Fig.\ref{fig:adv_train_img}.  The adversarial examples are similar to the clean images but the distortions are extremely obvious.

Additionally, when attackers attempt to reconstruct training data from adversarially-trained victim models, these corresponding confidence vectors of the training data (i.e., the classifier's output) will be used as inputs in the reconstruction process. Unlike the confidence vectors of adversarial examples, these confidence vectors of training data are normal and do not contain any perturbation information. Therefore, the reconstructed images against adversarially-trained victim models are also normal images instead of adversarial samples. We also conducted experiments by classifying the reconstructed images of MNIST dataset using the normally-trained classifier and the adversarially-trained classifier, which are shown in Table~\ref{tbl:adv_trained}. The majority of these reconstructed images are correctly classified by both the normally-trained and adversarially-trained classifiers. These results show that our method can also attack the adversarially-trained models.
\begin{table}
    \renewcommand{\arraystretch}{1.3}
    \caption{Attack performance against the adversarially-trained victim model. $\epsilon$ means the perturbation size in the training of the victim model. `Accuracy' refers to the classification accuracy on original clean images. `Normally' represents the classification accuracy of the normally-trained classifier on these reconstructed images, whereas `Adversarially' denotes the classification accuracy of the adversarially-trained classifier on these reconstructed images.}
    \label{tbl:adv_trained}
    \centering
    \begin{tabular}{c|l|c|cc}
         \hline
         $\epsilon$ &Accuracy& Method & Normally & Adversarially\\
         \hline
         \multirow{2}{*}{$0.02$} &\multirow{2}{*}{$99.7\%$}& baseline&$83.9\%$ & $81.0\%$\\
         & &our & $97.7\%$ & $93.5\%$\\
         \hline
         \multirow{2}{*}{$0.3$} &\multirow{2}{*}{$99.7\%$} & baseline & $93.2\%$ & $94.5\%$ \\
        & & our&$98.1\%$ & $\textbf{99.6\%}$ \\
         \hline
    \end{tabular}
\end{table}

\subsubsection{Other metrics: PSNR and SSIM}
\begin{figure}
    \centering
    \includegraphics[width=0.9\columnwidth]{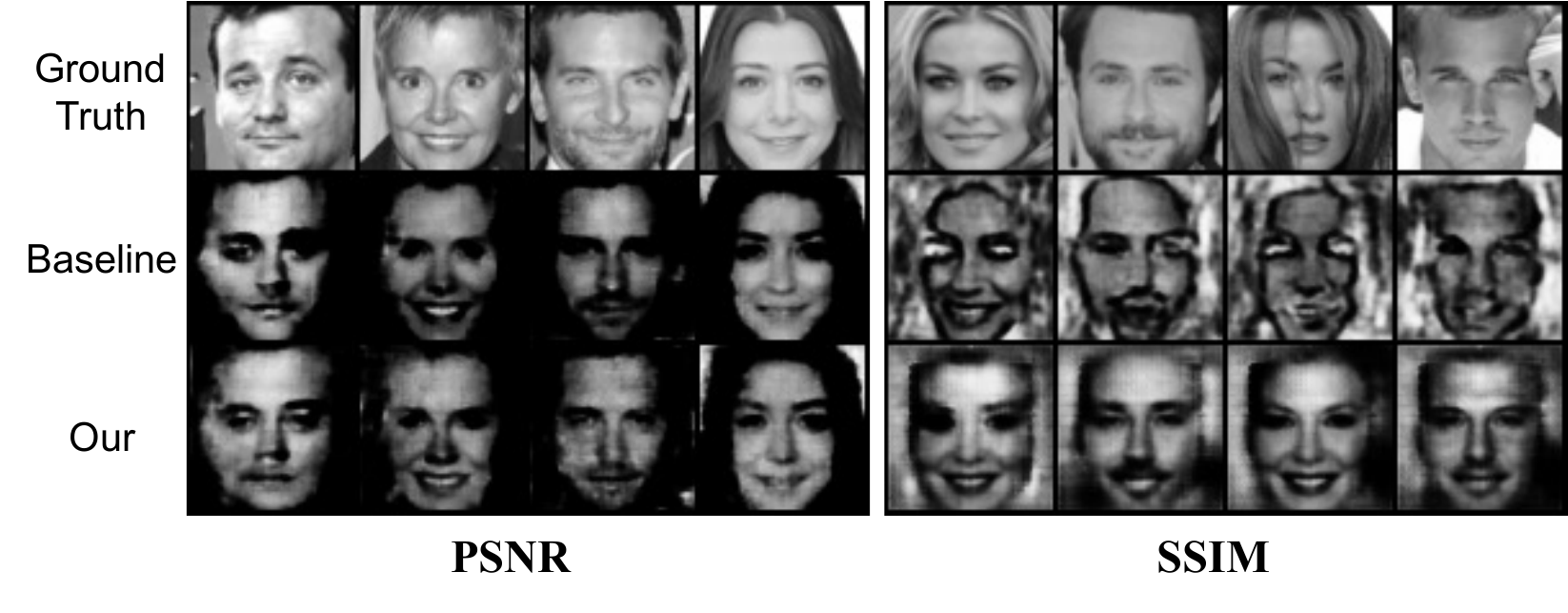}
    \caption{Comparison of reconstructed images using PSNR loss and SSIM loss. Our method applied a semantic loss with $\lambda = 0.002$.}
    \label{fig:psnr}
\end{figure}

Previous experiments have demonstrated that the MSE loss is an insufficient metric for model inversion attacks. In this study, we also conducted experiments to determine the efficacy of other metrics, Peak Signal to Noise Ratio (PSNR) and Structure Similarity (SSIM). The visual results were depicted in Fig.~\ref{fig:psnr}, while the quantitative outcomes were presented in Table~\ref{tab:psnr}. Our findings indicate that when we substitute MSE with PSNR and SSIM, the inversed results are comparable. We used three evaluation models with various architectures to calculate the attack accuracy. Similar to MSE loss, only PSNR or SSIM without our semantic loss is ineffective when training inversion models, as shown by these results. This is due to the fact that it is challenging to evaluate the quality of reconstructed images through only a simple metric (such as MSE, PSNR or SSIM). In contrast, we applied a neural network with feature extractors to evaluate the quality of the reconstructed images, and the output of this neural network is used to calculate the semantic loss, which can further enhance the attack accuracy from a machine's perspective.

\begin{table}
    \renewcommand{\arraystretch}{1.3}
    \caption{Attack performance comparison with PSNR and SSIM loss on the FaceScrub dataset.} 
    \label{tab:psnr}
    \centering
    \begin{tabular}{l|l|cccc}
         \hline
        \multirow{2}{*}{\textbf{Metrics}}&\multirow{2}{*}{\textbf{Methods}}&\multirow{2}{*}{\textbf{Loss}}&\multicolumn{3}{c}{\textbf{Evaluation Model}}\\
          &&&4-CNNs&ResNet&VGG\\
         \hline
         \multirow{2}{*}{PSNR}&baseline&$55.03$&$14.75\%$&$10.45\%$&$11.94\%$ \\
         &ours& $54.84$&$\textbf{23.40}\%$ &$\textbf{19.45}\%$ &$\textbf{16.23}\%$  \\
         \hline
         \multirow{2}{*}{SSIM}&baseline&$0.563$&$8.97\%$&$13.30\%$&$8.78\%$\\
          &ours&$0.564$& $\textbf{17.09}\%$&$\textbf{14.09}\%$&$\textbf{15.48}\%$  \\
         \hline
    \end{tabular}
\end{table}

\subsubsection{Training Time}
In addition, we examined the training time consumption of various methods. Our experiments were conducted on a workstation with a single NVIDIA GeForce RTX $3090$-$24$GB GPU. We have listed the average training time of the inversion model and the standard deviation (in the form of $a\pm b$) in Table~\ref{tbl:time}. Since our method requires a large number of adversarial examples prior to training the inversion model, the construction of these adversarial examples is a time-consuming process. The generation time is proportional to the number of iterations required during the generation procedure. After obtaining these adversarial examples, our inversion model's training time is longer than that of the baseline method due to the increased amount of training data. Nevertheless, it is worth noting that the increase in training time is acceptable as the inversion model requires only a single training session. Once the model is trained, the subsequent attack procedure involves only the forward inference, which is extremely fast and comparable to that of the baseline method. Thus, our method that effectively improves the attack accuracy of the inversion model, albeit at the cost of the increased training time, is worthwhile.
\begin{table}
    \renewcommand{\arraystretch}{1.3}
    \caption{Comparison of the training time of different inversion models. `Iterations' represents the number of iterations to generate adversarial examples, whereas `Generation' denotes the time required to produce adversarial examples. And `Training' refers to the time required to train inversion models.}
    \label{tbl:time}
    \centering
    \begin{tabular}{l|l|ccc}
         \hline
         \textbf{Dataset} & Methods & Iterations & Generation(s) & Training(s)\\
         \hline
        \multirow{3}{*}{MNIST}&Baseline&-&-&$729\pm 15$\\
        \cline{2-5}
          &\multirow{2}{*}{Our}&$100$&$219\pm3$&$1364\pm 2$ \\
          &&$500$&$1208\pm 73$&$1370\pm 7$\\
         \hline
         \multirow{3}{*}{FaceScrub}&Baseline&-&-&$563\pm2$ \\
         \cline{2-5}
         &\multirow{2}{*}{Ours}&$200$&$332\pm26$&$1385\pm4$\\
         && $500$& $692\pm4$&$1390\pm1$   \\
         \hline
    \end{tabular}
\end{table}

\section{Discussion of Defense Methods}
Our attack method aims to address the poor attack accuracy of earlier model inversion attacks, and improves upon previous methods by employing our adversarial augmentation technique. Existing defense strategies against model inversion attacks tend to decrease the dependency between inputs and outputs during classifiers' training~\cite{MID2021,bilateral2022,ye2022one,zhu2020more}. Nevertheless, our attack method can proactively explore and construct the relationship between inputs and outputs by generating adversarial examples against the target model. These known defensive strategies would be ineffective against our attack method. A potential defense method against our adversarial augmentation might be to increase the robustness of the target model, for example, by adversarial training techniques~\cite{MadryMSTV18,advanceAT}. In particular, some adversarial examples can be crafted beforehand and injected into the original training dataset, making the trained target model more resistant to adversarial attacks and increasing the complexity of searching adversarial examples against the trained target model. Consequently, our strategy for locating class-related parts in images based on adversarial examples may be weakened.

Methods to improve the robustness of models have been extensively investigated in the model security field~\cite{MadryMSTV18,advanceAT,FGSM}. However, it is important to note that these techniques may have unexpected side effects on privacy leakage issues, despite enhancing the target model's resistance to our adversarial augmentation. For example, Song et al.~\cite{song2019privacy} suggested that defensive strategies against adversarial attacks in the model security domain can indeed raise the probability of membership inference attacks against the target model. 

\begin{figure}[t]
\centering
\includegraphics[width=0.9\columnwidth]{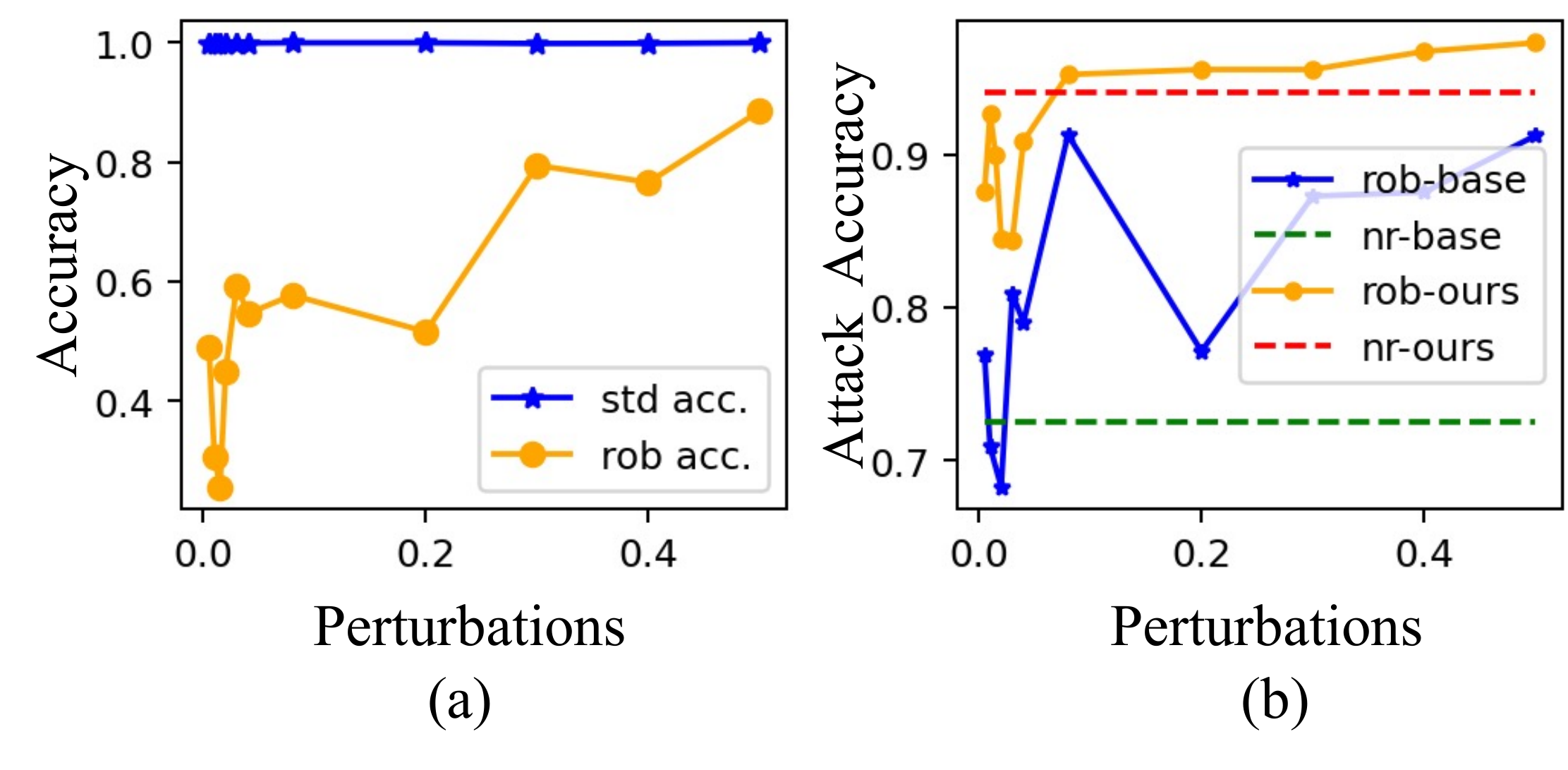} 
\caption{The classification accuracy of classifiers trained with standard training and with adversarial training, as well as model inversion attack accuracy against them. In (a), the label `std acc.' is an abbreviation for standard accuracy, and `rob acc.' represents robust accuracy. In (b), we compare the performance of baseline method on defended robust model (`rob-base') and undefended standard model (`nr-base'), as well as the performance of our attack method on defended robust model (`rob-ours') and undefended standard model (`nr-ours'). } 
\label{fig:def_curve}
\end{figure}

\begin{figure}[t]
\centering
\includegraphics[width=0.9\columnwidth]{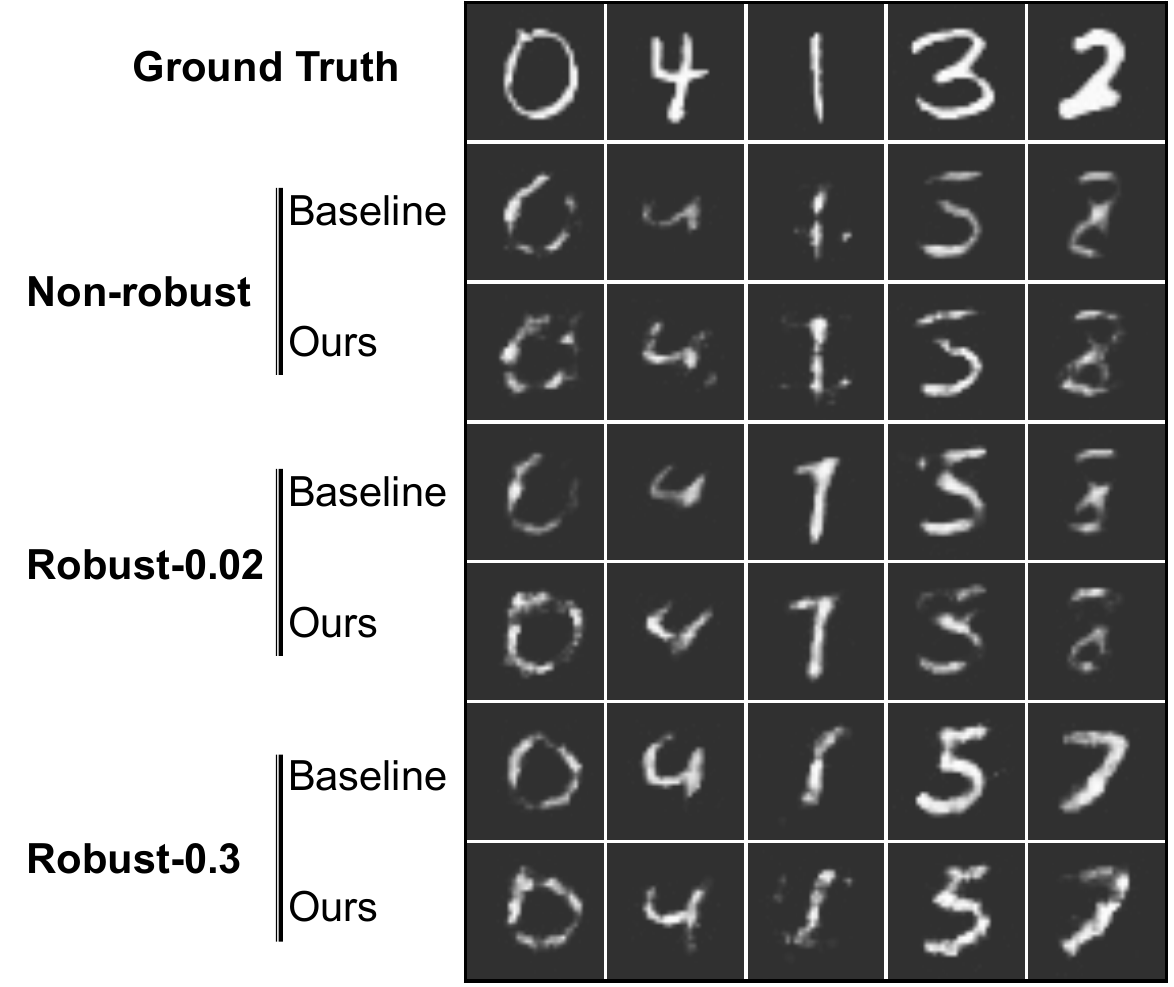} 
\caption{Comparison between reconstructed images against undefended and defended classifiers. `Non-robust' represents undefended classifier with standard training procedure. `Robust-$\gamma$' indicates the robust classifier with adversarial training strategy, where $\gamma$ is the bound of adversarial perturbations used in the adversarial training.} 
\label{fig:defense}
\end{figure}

We also applied this potential defense strategy, adversarial training~\cite{MadryMSTV18,advanceAT}, in our experiments on MNIST, where the FGSM method~\cite{FGSM} is used to generate adversarial examples. Fig.~\ref{fig:def_curve} displays the experimental results. We plot the standard accuracy (i.e., accuracy on clean examples) and robust accuracy (i.e., accuracy on adversarial examples) with respect to the bound of perturbation used in the adversarial training. This shows that the adversarial training strategy has a small effect on the standard accuracy of the classifier on the MNIST dataset, whereas the robustness can be significantly enhanced. When the perturbation bound used in adversarial training is small, adversarial training can reduce the attack accuracy of the baseline technique and our attack method. However, as the perturbation bound expands, adversarial training may increase the privacy leakage. This occurrence is consistent with Song et al.'s research findings~\cite{song2019privacy}. Fig.~\ref{fig:defense} also depicts the visual results. When the classifier is adversarially trained with the perturbations bounded by $0.02$, the reconstruction performance of the baseline method and our attack method are influenced. For instance, the reconstructed images of digit $1$ resemble digit $7$. Namely, the privacy protection of the classifier is improved. Nonetheless, when the perturbation in adversarial training is increased to $0.3$, the reconstructed images of digit $2$ against the robust classifier have a higher degree of resemblance with the ground truth than the reconstructions against the non-robust classifier. 

This unexpected side effect of adversarial training may be due to the fact that the robustness improvement diminishes the target classifier's sensitivity to noises. The classifier's overfitting is reduced. As a result, it may be easier to train a shadow classifier with a higher degree of similarity to the target classifier, allowing the training of powerful inversion models. Consequently, the efficacy of adversarial defence in defending our attack must be thoroughly investigated, which will constitute our future research in this field.

\section{Conclusion}
In this paper, we proposed a novel learning-based model inversion attack that shows increased attack accuracy and has a better capacity for reproducing semantics than the current black-box benchmark. Two factors contribute to the increased performance of the attack. First, the training loss function of the inversion model introduces a penalty term for semantic reproduction loss. Second, an adversarial augmentation technique improves the diversity of the class-related information in the training data. The attack's accuracy is improved to such an extent that the reconstructed images can be recognized by both humans and machines. Extensive experiments show the effectiveness and efficiency of the attack. The work presented should motivate the field to rethink the severity of the threat that learning-based model inversion attacks present. Further, the results should provide support for a more accurate assessment of the privacy and security concerns of machine learning models.

\section*{Acknowledgment}
This paper is supported by the Australian Research Council Discovery DP$200100946$ and DP$230100246$. 


\ifCLASSOPTIONcaptionsoff
  \newpage
\fi

\bibliographystyle{IEEEtran}
\bibliography{TDSC}

\end{document}